\documentstyle[12pt]{article}

\newcommand{\tr}{{\rm tr}}
\newcommand{\Tr}{{\rm Tr}}

\newcommand{\be}{\begin{eqnarray}}
\newcommand{\ee}{\end{eqnarray}} 

\begin{document} 

\begin{titlepage}

\hrule 
\leftline{}
\leftline{Preprint
          \hfill   \hbox{\bf CHIBA-EP-99-REV}}
\leftline{\hfill   \hbox{hep-th/9709109(revised)}}
\leftline{\hfill   \hbox{September 1997}}
\vskip 5pt
\hrule 
\vskip 1.0cm

\centerline{\large\bf 
Abelian-Projected Effective Gauge Theory of QCD 
} 
\centerline{\large\bf  
{
with Asymptotic Freedom and Quark Confinement 
}
}
   
\vskip 0.5cm

\centerline{{\bf 
Kei-Ichi Kondo$^{1}{}^{\dagger}$
}}  
\vskip 4mm
\begin{description}
\item[]{\it  
$^1$ Department of Physics, Faculty of Science,
  Chiba University, Chiba 263, Japan
  }
\item[]{$^\dagger$ 
  E-mail:  kondo@cuphd.nd.chiba-u.ac.jp 
  }
\end{description}
\vskip 0.5cm

\centerline{{\bf Abstract}} \vskip .5cm
Starting from SU(2) Yang-Mills theory in 3+1 dimensions,
we prove that the abelian-projected effective gauge
theories are written in terms of the maximal abelian gauge
field and the dual abelian gauge field interacting with
monopole current.  This is performed by integrating out
all the remaining non-Abelian gauge field belonging to
SU(2)/U(1).  We show that the resulting abelian gauge
theory recovers exactly the same one-loop beta function as
the original Yang-Mills theory.  Moreover, the dual abelian
gauge field becomes massive if the monopole condensation
occurs.  This result supports the dual superconductor
scenario for quark confinement in QCD.  We give a
criterion of dual superconductivity and point out that
the monopole condensation can be estimated from the
classical instanton configuration. Therefore there can
exist the effective abelian gauge theory which shows both
asymptotic freedom and quark confinement based on the dual
Meissner mechanism. Inclusion of arbitrary number of
fermion flavors is straightforward in this approach. Some
implications to lower dimensional case will also be
discussed.

\vskip 0.5cm

\vskip 0.5cm
\hrule  

\vskip 0.5cm  


\end{titlepage}

\pagenumbering{arabic}

\newpage
\section{Introduction}
\setcounter{equation}{0}
\par

It is one of the most important problems in particle
physics to clarify the physical mechanism which realizes
the quark and gluon confinement.
An important question is what is the most relevant degrees
of freedom to describe the confinement.
In the mid-1970's, an idea of the dual Meissner vacuum of
quantum chromodynamics (QCD) was proposed by Nambu
\cite{Nambu74}, 't Hooft \cite{tHooft81} and Mandelstam
\cite{Mandelstam76}.
In this scenario, the monopole degrees of freedom plays
the most important role in the confinement.
This aspect can be seen explicitly through a procedure
called the {\it abelian projection} by 't Hooft
\cite{tHooft81}. Under the abelian projection, the
non-Abelian gauge theory can be regarded as an abelian
gauge theory with magnetic monopole \cite{Dirac31}.  
For the confinement mechanism, there are other proposals
\cite{tHooft78} which we do not discuss in
this paper.
\par
The abelian projection \cite{tHooft81} is to fix the gauge
in such a way that the maximal torus group of the
gauge group $G$ remains unbroken. 
It goes on as follows  for the gauge group SU(N),
\begin{enumerate}
\item[1)]
 One chooses a
gauge-dependent local quantity 
$X(x)=X^A(x)T^A$ which transforms adjointly under the gauge
transformation, i.e.,
\begin{eqnarray}
   X(x) \rightarrow X'(x) := U(x) X(x) U^\dagger(x) .
\end{eqnarray}
\item[2)]
 One performs the gauge rotation so that $X$ becomes
diagonal, 
\begin{eqnarray}
    X'(x) = diag(\lambda_1(x), \cdots , \lambda_N(x)) ,
\end{eqnarray}
where $\lambda_i(x) \ (i=1, \cdots , N)$ are eigenvalues.
\item[3)]
At the space-time point where the eigenvalues are
degenerate
$
 \lambda_i(x)=\lambda_j(x) (i\not=j, i,j = 1, \cdots, N) ,
$
the monopole-like (hedgehog) singularity appears.
The singularity does appear in the abelian gauge field
$a_\mu(x)$ extracted from the non-Abelian gauge field
${\cal A}_\mu'(x)= U(x)({\cal A}_\mu(x)+
{i \over g}\partial_\mu)U^\dagger(x)$. 
The monopole singularity is
characterized  as a topological quantity.
\item[4)]
 At generic point where the eigenvalues do not coincide,
the gauge is not determined completely, since any diagonal
gauge rotation $U$ (an element of the largest abelian
subgroup, $U(1)^{N-1}$, the maximal torus group)
\begin{eqnarray}
    U(x) = diag(e^{i\theta_1(x)}, \cdots ,
e^{i\theta_N(x)}) , \quad
\sum_{i=1}^N \theta_i(x) = 0,
\end{eqnarray}
leaves $X$ invariant.  Therefore, within this gauge, the
theory reduces to an (N-1) fold abelian gauge-invariant
theory.
\end{enumerate}

\par
The Monte Carlo studies of the abelian projection was
initiated by the work \cite{KSW87} and the maximal abelian
gauge (MAG) was adopted in the simulation on the
lattice \cite{KLSW87}.   
Recent extensive studies of abelian projection (see
\cite{Polikarpov96} for a review) have confirmed the {\it
abelian dominance} proposed in Ref.\cite{EI82}.   This
states that the non-Abelian gauge field $A_\mu^a$ in
$SU(N)/U(1)^{N-1}$, behaving as a charged field under the
residual
$U(1)^{N-1}$ gauge rotation, is not important in the low
energy physics and the maximal abelian part
$U(1)^{N-1}$ plays the dominant role in the quark and gluon
confinement.  
In analytical studies, the abelian dominance was assumed
from the beginning to construct the effective low
energy theory of QCD
\cite{EI82,Suzuki88}.
Assuming the abelian dominance, one
can show that, if the monopole condensation occurs, charged
quarks and gluons are confined due to dual Meissner effect.
The monopole condensation is expected to bring the mass
for the dual gauge field.
An effective theory of monopole currents was investigated
also on the lattice \cite{SS91}.
In fact, recent Monte Carlo simulations 
\cite{SY90} support the abelian dominance and furthermore
{\it monopole dominance}. However, there seems to be no
analytical proof of abelian dominance.  
\par
An deficit of the abelian projection is the
gauge-dependence of the procedure of abelian projection.
The quantity $X$ is a gauge-dependent quantity and the
field variable in which the monopole appears is not a
gauge invariant quantity.   Therefore the result seems to
depend crucially on the gauge selected in the abelian
projection.  However, this would not be a real problem,
since it is possible to put the abelian
projection in a gauge-invariant form,  if we desire to do
so, see
\cite{BOT96,Shabanov97}.
\par
The real problem is another in our view.
In the abelian-projected theory, the magnetic monopole
degrees of freedom appear as the singularity in the
abelian gauge field.   
The magnetic current $k_\mu$ is obtained as
the divergence of the dual  abelian field strength $\tilde
f_{\mu\nu}$,
\begin{eqnarray}
  \partial^\nu \tilde f_{\mu\nu} = k_\mu, 
  \quad 
 \tilde f_{\mu\nu} := {1 \over 2} 
\epsilon_{\mu\nu\rho\sigma} f^{\rho\sigma} ,
\end{eqnarray}
in the similar way that the equation of motion 
relates the field strength $f_{\mu\nu}$ to the electric
current
$j_\mu$,
\begin{eqnarray}
  \partial^\nu  f_{\mu\nu} = j_\mu .
\end{eqnarray}
If the U(1) potential $a_\mu$ is non-singular, the
abelian field strength 
$f_{\mu\nu}:=\partial_\mu a_\nu - \partial_\nu a_\mu$ leads
to vanishing magnetic current, $k_\mu=0$, which is
nothing but the Bianchi identity for the U(1) field,
$\partial^\nu \tilde f_{\mu\nu} \equiv 0$. So,
if one needs the non-zero magnetic current, the abelian
field must include a singularity.  
However, we do not think that it is sound as a
quantum field theory to treat the singularity of the field
variable as the essential ingredient  from the very
outset.  In the lattice gauge theory, such a singularity
does not appear due to the lattice regularization
\cite{DT80} and the monopole contribution is extracted
from the gauge-invariant magnetic flux, although the
monopole dominance is supported in the Monte Carlo
simulation on the lattice. Moreover, it should be noted
that the magnetic monopole does not exists in
the original non-Abelian gauge theory. Magnetic monopole
appears only after the abelian projection (see Appendix
C).   
\par
The purpose of this paper is to {\it derive} the 
abelian-projected effective gauge theory (APEGT) of QCD
{\it as a quantum field theory}, from which we should start
the analysis. 
For simplicity, we restrict the following argument to the
$G=SU(2)$ case. $SU(3)$ case is more involved and will be
presented in a subsequent paper. 
In this paper, without using various assumptions
(actually with no assumptions), we derive the  
APEGT of Yang-Mills (YM) theory and QCD. 
This is done by integrating out  off-diagonal
fields belonging to the SU(2)/U(1) based on the functional
integral formalism.  
We use the word "effective" in the sense of the Wilson
renormalization group \cite{Wilson}, since the
abelian-projected theory is obtained after integrating out
the  degrees of freedom corresponding to non-abelian gauge
fields
$
 A_\mu^{\pm} := (A_\mu^1 \pm i A_\mu^2)/\sqrt{2}
$ 
which behave as {\it massive} charged matter fields and
don't play the important role in the low energy physics of
confinement.  Such a strategy can be exactly performed in
the $N=2$ supersymmetric YM and QCD \cite{SW94}.
\par
We show that the off-diagonal field 
gives rise to the non-trivial magnetic monopole current
for the abelian part,
\begin{eqnarray}
  K^\mu = {1 \over 2} \partial_\nu  (
\epsilon^{\mu\nu\rho\sigma} \epsilon^{ab3} A_\rho^a
A_\sigma^b) ,
\quad a, b = 1,2.
\label{dmc}
\end{eqnarray}
In other words, the charged off-diagonal gluon field
plays the role of the source of the monopole.  
Although the definition (\ref{dmc}) of monopole current
seems to be different from the usual definition based on
the singularity of the abelian field, we show that both are
equivalent to each other (apart from the Dirac string
singularity). In the APEGT, the singularity does not appear
apparently, although we can always include the singularity
if necessary.
\par
The effective dual Ginzburg-Landau (GL) theory derived
assuming the abelian dominance does not have sufficient
predictive power, since it contains undetermined free
parameters. 
On the contrary, all the quantities in APEGT are
calculable and all the effects of the non-Abelian gauge
field are included in the APEGT.
In fact, we show that the APEGT recovers
exactly the same one-loop beta function as that of the
original non-Abelian gauge theory.
The dual abelian gauge field follows naturally in the
course of the derivation of the theory and has a coupling
with the monopole current.  This interaction leads to the
dual Meissner effect due to monopole condensation. The
resulting non-zero mass of the dual gauge field gives the
non-zero string tension, i.e. linear potential for
static quarks. Thus the string tension is determined by the
monopole loop condensate,
$
 \langle K_\mu(x) K^\mu(x) \rangle/\delta^{(4)}(0) ,
$
(see section 4 for precise definition).
The monopole condensate plays
the role of the order parameter for confinement. 
\par
Moreover, we discuss a possibility that the non-zero
monopole condensation is derived from the instanton
configuration. Hence instanton may lead to the confinement
against the conventional wisdom
\cite{CDG78}.
\par
In our approach, the inclusion of fermions
is straightforward.   Hence APEGT
 is also a starting point to study the relationship
between the confinement and the chiral symmetry breaking
(or restoration)
\cite{SST95,Miyamura95}. 
\par

\par
This paper is organized as follows.
In section 2, we derive the APEGT
for the maximal abelian part by integrating out the
remaining non-Abelian gauge field.  
In this step, we introduce the auxiliary tensor field which
is converted to the dual gauge field.
The dual gauge field is essential to discuss the
dual Meissner effect in section 4.
APEGT is first obtained in the form including the
logarithmic determinant. The logarithmic determinant is
explicitly calculated.  It generates the gauge invariant
form due to U(1) gauge invariance. An effect of this term
is the renormalization of the abelian gauge field.
In section 3, we calculate the one-loop beta function
without using the Feynman diagram.  It is shown to agree
with the original non-Abelian gauge theory.  In this
sense, the effective theory recovers the asymptotic freedom.
In section 4, we discuss the dual Meissner effect.  If the
monopole loop condensation occurs, the dual vector field
becomes massive.
In section 5, we include the fermion into the APEGT.
In section 6, we discuss the lower dimensional case.
In the final section we give conclusion and discussion.

\section{Abelian-projected effective gauge theory}
\setcounter{equation}{0}

\par
\subsection{Separation of the abelian part and introduction
of the dual field}
 First, we decompose the field $ {\cal A}_\mu $
into the diagonal (maximal abelian U(1)) and the
off-diagonal part SU(2)/U(1),
\begin{eqnarray}
 {\cal A}_\mu(x) = \sum_{A=1}^3  {\cal A}_\mu^A(x) T^A
 :=  a_\mu(x)  T^3 
 + \sum_{a=1}^{2}  A_\mu^a(x) T^a .
\end{eqnarray}
We adopt the following convention.
The generators of the Lie algebra $T^A(A=1, \cdots,
N^2-1)$ for the gauge group $G=SU(N)$ are taken to be
hermitian satisfying 
$
 [T^A, T^B] = i f^{ABC} T^C
$
and normalized as
$
 \tr(T^A T^B) =  {1 \over 2} \delta^{AB}.
$
The generators in the adjoint representation is given by
$
 [T^A]_{BC} = -i f_{ABC} .
$
We define the quadratic Casimir by
$
 C_2(G) \delta^{AB} = f^{ACD}f^{BCD}.
$
For SU(2), $T^A = (1/2) \sigma^A (A=1,2,3)$ with Pauli
matrices $\sigma^A$ and the structure constant
$f^{ABC} = \epsilon^{ABC}$.
The indices $a, b, \cdots$ denote the off-diagonal parts.
\par
Then the field strength 
\begin{eqnarray}
 {\cal F}_{\mu\nu}(x) 
:= \sum_{A=1}^{3} {\cal F}_{\mu\nu}^A(x) T^A
:=   \partial_\mu {\cal A}_\nu(x) 
 -   \partial_\nu {\cal A}_\mu(x)
 - i [{\cal A}_\mu(x), {\cal A}_\nu(x)]
\end{eqnarray}
is decomposed as 
\begin{eqnarray}
 {\cal F}_{\mu\nu}(x) 
 &=&  [f_{\mu\nu}(x) + {\cal C}_{\mu\nu}(x)]T^3 
 + {\cal S}_{\mu\nu}^a(x)T^a ,
\nonumber\\
 f_{\mu\nu}(x) &:=& \partial_\mu a_\nu(x) 
 -   \partial_\nu a_\mu(x) ,
\nonumber\\
 {\cal S}_{\mu\nu}^a(x) &:=& 
 D_\mu[a]^{ab} A_\nu^b - D_\nu[a]^{ab}A_\mu^b ,
\nonumber\\
 {\cal C}_{\mu\nu}(x)T^3 &:=& - i[ A_\mu(x), A_\nu(x)] ,
\end{eqnarray}
where the derivative $D_\mu[a]$ is defined by
\begin{eqnarray}
 D_\mu[a]  = \partial_\mu   + i[a_\mu T^3, \cdot ~],
\quad
 D_\mu[a]^{ab} := \partial_\mu \delta^{ab}
 - \epsilon^{ab3} a_\mu .
\end{eqnarray}
Hence the diagonal part ${\cal F}_{\mu\nu}^3$ of the field
strength is given by
\begin{eqnarray}
 {\cal F}_{\mu\nu}^3 = f_{\mu\nu} + {\cal C}_{\mu\nu},
 \quad 
 {\cal C}_{\mu\nu} := \epsilon^{ab3}A_\mu^a A_\nu^b .
\end{eqnarray}

\par
Next, we rewrite the Yang-Mills (YM) action
\begin{eqnarray}
   S_{YM}[{\cal A}] = -{1 \over 2g^2} \int d^4x \
   \tr({\cal F}_{\mu\nu} {\cal F}^{\mu\nu}) .
\end{eqnarray}
By using 
\begin{eqnarray}
 \tr(f_{\mu\nu}{\cal S}^{\mu\nu}) = 0 
 = \tr({\cal C}_{\mu\nu}{\cal S}^{\mu\nu}) ,
\end{eqnarray}
the YM action is rewritten as
\begin{eqnarray}
   S_{YM}[{\cal A}] = -{1 \over 4g^2} \int d^4x \
   [(f_{\mu\nu} + {\cal C}_{\mu\nu})^2
    + ({\cal S}_{\mu\nu}^a)^2] .
\label{YM2}
\end{eqnarray}
\par
Here we introduce an antisymmetric auxiliary tensor field
$B_{\mu\nu}$ in order to linearize the 
$({\cal C}_{\mu\nu})^2$ term.  This procedure enables us
to perform the Gaussian integration over the off-diagonal
gluon fields $A_\mu^a (a=1,2)$. 
\footnote{
This procedure is similar to the field strength approach
for non-Abelian gauge theory
\cite{Halpern77}.
} 
It turns out that the tensor field $B_{\mu\nu}$ plays the
role of the "dual" field to the abelian gluon field
$a_\mu$.  We find that there are two ways to introduce the
"dual" tensor field.  
\par
One way is to introduce the tensor field
$B_{\mu\nu}$ such that the tensor $B_{\mu\nu}$ is the dual
of the diagonal field strength  ${\cal F}^3_{\rho\sigma}$,
\begin{eqnarray}
  B_{\mu\nu}  \leftrightarrow {1 \over 2}
\epsilon^{\mu\nu\rho\sigma} {\cal F}^3_{\rho\sigma}
=  {1 \over 2} \epsilon^{\mu\nu\rho\sigma}
(f_{\rho\sigma} + {\cal C}_{\rho\sigma}) .
\end{eqnarray}
This is achieved in the tree level by the 
following action
\begin{eqnarray}
   S_{apBFYM}[{\cal A}, B] =  \int d^4x \left[
     {1 \over 4} \epsilon^{\mu\nu\rho\sigma} B_{\rho\sigma} 
     (f_{\mu\nu} + {\cal C}_{\mu\nu}) 
     - {1 \over 4} g^2 B_{\mu\nu} B^{\mu\nu} 
    - {1 \over 4g^2} ({\cal S}_{\mu\nu}^a)^2 \right] .
    \label{apBFYM}
\end{eqnarray}
This theory is equivalent to the BF-YM theory,
\begin{eqnarray}
   S_{BFYM}[{\cal A}, {\cal B}] =  \int d^4x \left[
     {1 \over 4} \epsilon^{\mu\nu\rho\sigma}
{\cal B}_{\rho\sigma}^A   {\cal F}_{\mu\nu}^A 
  - {1 \over 4}  g^2 {\cal B}_{\mu\nu}^A 
  {\cal B}^{\mu\nu}{}^A
\right] .
\label{BFYM}
\end{eqnarray}
Actually, by identifying 
$B_{\mu\nu}={\cal B}_{\mu\nu}^3$,
the action (\ref{apBFYM}) is obtained from (\ref{BFYM})
by separating the diagonal part from the off-diagonal part
and integrating out the off-diagonal auxiliary tensor field
$B_{\mu\nu}^a(a=1,2)$. Quite recently, equivalence of the
BF-YM theory with the YM theory has been proved in the
quantum level, see
\cite{BFYM}.  
This theory is interesting from  the 
topological point of view.
\par
Another way is to introduce the tensor field as a dual to 
${\cal C}_{\rho\sigma}$ at the tree level,
\begin{eqnarray}
  B_{\mu\nu}  \leftrightarrow   {1 \over 2}
\epsilon^{\mu\nu\rho\sigma} {\cal C}_{\rho\sigma} .
\end{eqnarray}
Thus we are lead to the action,
\begin{eqnarray}
   S_{apYM}[{\cal A}, B] &=&  \int d^4x \Big[
     - {1 \over 4g^2} 
     (f_{\mu\nu}f_{\mu\nu} + 2 f_{\mu\nu}{\cal C}_{\mu\nu}) 
  +  {1 \over 4} \epsilon^{\mu\nu\rho\sigma} B_{\rho\sigma} 
   {\cal C}_{\mu\nu}
     - {1 \over 4} g^2 B_{\mu\nu} B^{\mu\nu} 
\nonumber\\&&
    - {1 \over 4g^2} ({\cal S}_{\mu\nu}^a)^2 \Big] .
    \label{apYM}
\end{eqnarray}
In this case, 
$
{1 \over 2}
\epsilon^{\mu\nu\rho\sigma} f_{\rho\sigma}
$
is generated through the radiative correction as shown in
 section 2.4. In either case, Gaussian integration over
$B_{\mu\nu}$ recovers the action (\ref{YM2}) and hence the
original YM action.   This model (\ref{apYM}) is simpler
than the model (\ref{apBFYM}) in the actual treatment,
since the topological theory need some delicate treatment
\cite{BFYM}. (Equivalence of two formulations is shown in
Appendix A.)
In what follows, we focus on the action
(\ref{apYM}) which is essentially equivalent to that
derived by Quandt and Reinhardt
\cite{QR97}.

\subsection{Gauge-fixing}

We discuss the gauge-fixing term.  This is independent from
the choice of the action.
 The gauge-fixing term is constructed
based on the BRST formalism.
  We consider the gauge given by
\begin{eqnarray}
 F^{\pm}[A,a] &:=& (\partial^\mu \pm i \xi a^\mu)
A_\mu^{\pm} = 0,
\label{dMAG}
\\
  F^3[a] &:=& \partial^\mu  a_\mu = 0 ,
 \label{GF}
\end{eqnarray}
where we have used the $(\pm, 3)$ basis,
\footnote{
In this basis, 
$
 \sum_{\pm} P^{\pm}Q^{\mp} = P^+ Q^- + P^- Q^+ = P^a Q^a ,
 \quad
 \sum_{\pm} (\pm) P^{\mp}Q^{\pm} = - P^+ Q^- + P^- Q^+ =  i
\epsilon^{ab3} P^a Q^b (a,b=1,2).
$
}
\begin{eqnarray}
{\cal O}^{\pm} := ({\cal O}^1 \pm i{\cal O}^2)/\sqrt{2} .
\end{eqnarray}
The gauge fixing with $\xi=0$ is the Lorentz gauge,
$\partial^\mu {\cal A}^\mu = 0$.
In particular, $\xi=1$ corresponds to the differential form
of the maximal abelian gauge (MAG) which is expressed as
the minimization of the functional 
\begin{eqnarray}
  {\cal R}[A] := {1 \over 2}\int d^4x [(A_\mu^1(x))^2 +
(A_\mu^2(x))^2 ]
  = \int d^4x A_\mu^{+}(x) A_\mu^{-}(x)   .
  \label{MAG}
\end{eqnarray}
The differential MAG condition (\ref{dMAG}) corresponds to
 a local minimum of the gauge fixing functional
$R[A]$, while the MAG condition (\ref{MAG}) requires the
global (absolute) minimum.  
The differential MAG condition (\ref{dMAG}) fixes gauge
degrees of freedom in SU(2)/U(1) and is invariant under the
residual U(1) gauge transformation.  An additional
condition (\ref{GF}) fixes the residual U(1) invariance. 
Both conditions (\ref{dMAG}) and 
(\ref{GF}) then completely fix the gauge except
possibly for the Gribov problem.
It is known that the differential MAG
(\ref{dMAG}) does not spoil renormalizability of YM theory 
\cite{MLP85}.  An implication of this fact is shown in
Appendix B.
\par
From physical point of view, we expect that MAG introduces
the non-zero mass $m_{A}$ for the off-diagonal gluons,
$A_\mu^1, A_\mu^2$.
This is suggested from the form (\ref{MAG}) which is
equal to the mass term for $A_\mu^1, A_\mu^2$, although
we need an independent proof of this statement.  This
motivates us to integrate out the off-diagonal gluons in
the sense of Wilsonian renormalization group and allows us
to regard the resulting theory as the low-energy effective
gauge theory written in terms of massless fields alone
which describes the physics in the length scale 
$R > m_{A}^{-1}$.
The abelian dominance will be realized in the physical
phenomena occurring in the scale $R > m_{A}^{-1}$. 
In this sense the choice of MAG is not unique in realizing
abelian dominance.  We can equally take the gauge so
that the off-diagonal gluon fields acquire non-zero
masses.  Then the abelian-projected  effective
gauge theory obtained by integrating out the massive
off-diagonal gluons will be valid in the low energy region
below the energy scale given by the off-diagonal gluon
mass.  
\par
We introduce the Lagrange multiplier field
$\phi^{\pm}$ and $\phi^3$ for the gauge-fixing
function $F^{\pm}[A]$ and
$F^{3}[A]$, respectively. 
It is well known that the gauge fixing term in the
BRST quantization is given by
\cite{KU82}
\begin{eqnarray}
  {\cal L}_{GF} = - i \delta_B G ,
    \label{BRSTGF}
\end{eqnarray}
where $G$ carries the ghost number $-1$ and is a hermitian
function of Lagrange multiplier field $\phi^{\pm}, \phi^3$,
ghost field $c^A$, antighost field $\bar c^A$, and the
remaining field variables of the original lagrangian.   In
this paper we consider a simple gauge given by
\begin{eqnarray}
  G = \sum_{\pm}
  \bar c^{\mp} (F^{\pm}[A,a] + {\alpha \over 2} \phi^{\pm})
  + \bar c^3 (F^{3}[a] + {\beta \over 2} \phi^{3}) .
\label{G1}
\end{eqnarray}
For the most general gauge fixing, see \cite{HN93}.
\par
The BRST transformation in the usual basis is
\begin{eqnarray}
   \delta_B {\cal A}_\mu  &=&  {\cal D}_\mu c 
   := \partial_\mu c - i [{\cal A}_\mu, c],
    \nonumber\\
   \delta_B c  &=&  i{1 \over 2} [c, c] ,
    \nonumber\\
   \delta_B \bar c  &=&   i \phi  ,
    \nonumber\\
   \delta_B \phi  &=&  0 ,
    \nonumber\\
   \delta_B {\cal B}_{\mu\nu} 
   &=&  -i[c, {\cal B}_{\mu\nu}] .
    \label{BRST0}
\end{eqnarray}
Then the BRST transformation 
in the $(\pm, 3)$ basis is given by
\begin{eqnarray}
   \delta_B A_\mu^{\pm} &=&  
  (\partial_\mu \pm i a_\mu) c^{\pm} \mp i A_\mu^{\pm} c^3,
    \nonumber\\
   \delta_B a_\mu  &=&   \partial_\mu c^3
   + i(A_\mu^{+} c^{-} - A_\mu^{-} c^{+}),
    \nonumber\\
   \delta_B c^{\pm} &=&   \mp i c^3 c^{\pm} ,
    \nonumber\\
   \delta_B c^3 &=&   - i c^{+} c^{-},
    \nonumber\\
   \delta_B \bar c^{\pm,3} &=&  i \phi^{\pm,3} ,
    \nonumber\\
   \delta_B \phi^{\pm,3} &=&  0 ,
    \nonumber\\
   \delta_B B_{\mu\nu}^{\pm} &=&  
   \mp i c^{\pm} B_{\mu\nu}^3  \pm i c^3 B_{\mu\nu}^{\pm} ,
    \nonumber\\
   \delta_B B_{\mu\nu}^3 &=&   
   i (c^{+} B_{\mu\nu}^{-} - c^{-} B_{\mu\nu}^{+}) .
    \label{BRST1}
\end{eqnarray}
Under the local U(1) gauge transformation,
\begin{eqnarray}
 a_\mu \rightarrow a_\mu + \partial_\mu \omega, 
 \quad
 {\cal O}^{\pm} \rightarrow e^{\mp i \omega} {\cal O}^{\pm}
 \quad
 {\cal O}^{3} \rightarrow {\cal O}^{3} .
\end{eqnarray}
Hence $a_\mu$ transforms as a U(1) gauge field, while
 $A_\mu^{\pm}$  and $B_{\mu\nu}^{\pm}$  behave as
charged matter fields under the U(1) gauge transformation.
It turns out that
$B_{\mu\nu}^3$ and 
\begin{eqnarray}
 {\cal C}_{\mu\nu} 
= i \sum_{\pm} (\pm) A_\mu^{\pm} A_\nu^{\mp}
\end{eqnarray}
are U(1) gauge invariant as expected. 
\par
In the usual basis, we can write
\begin{eqnarray}
  G = \sum_{a=1,2}
  \bar c^{a} (F^{a}[A,a] + {\alpha \over 2} \phi^{a})
  + \bar c^3 (F^{3}[a] + {\beta \over 2} \phi^{3}) ,
\label{G0}
\end{eqnarray}
where
\begin{eqnarray}
   F^{a}[A,a] := (\partial^\mu \delta^{ab} 
   - \xi \epsilon^{ab3} a^\mu) A_\mu^b 
   := D^\mu{}^{ab}{}[a]^\xi A_\mu^b.
\end{eqnarray}
\par
For the gauge fixing function (\ref{G1}) with the BRST
transformation (\ref{BRST1}), or (\ref{G0}) with
(\ref{BRST0}), straightforward calculation leads to the
gauge fixing lagrangian (\ref{BRSTGF}),
\begin{eqnarray}
  {\cal L}_{GF} &=& 
  \phi^a F^a[A,a] + {\alpha \over 2} (\phi^a)^2
  + i \bar c^a D^\mu{}^{ab}[a]^\xi D_\mu^{bc}[a] c^c
  + i \bar c^3 \partial^\mu \partial_\mu c^3
  \nonumber\\
  &&- i \xi \bar c^a [A_\mu^a A^\mu{}^b - A_\mu^c A^\mu{}^c
\delta^{ab} ] c^b
+ i \bar c^a \epsilon^{ab3} [(1-\xi) A_\mu^b \partial^\mu
  + F^b[A,a] ] c^3
  \nonumber\\
  && + \phi^3 F^3[a] + {\beta \over 2} (\phi^3)^2
  - i  \bar c^3 \partial^\mu (\epsilon^{ab3} A_\mu^a c^b) .
    \label{BRSTGF2}
\end{eqnarray}
This reduces to the usual form in the Lorentz gauge,
$\xi=0$.
\par
Finally we introduce the source term,
\begin{eqnarray}
  {\cal L}_{J} &=& 
  A_\mu^a J^\mu{}^a + \phi^a J_\phi^a ,
    \label{source}
\end{eqnarray}
which will be necessary to calculate the correlation
functions.

\subsection{Integration over SU(2)/U(1)}
\par
Our strategy is to integrate out the off-diagonal fields, 
$\phi^a, A_\mu^a, c^a, \bar c^a$ (and $B_{\mu\nu}^a$ for
BF-YM case) belonging to the Lie algebra of SU(2)/U(1) and
to obtain the effective abelian gauge theory written in
terms of the diagonal fields
$a_\mu, B_{\mu\nu}$ (and ghost fields $c^3, \bar c^3$ if
we need a completely gauge-fixed theory also for the
residual U(1) gauge invariance). 
\par
First of all, when
$\alpha \not= 0$,
\footnote{
The case of $\alpha=0$ should be treated separately.
Since $F^a[A,a]=DA$ is linear in $A_\mu^a$, the $\phi^a$
integration can be performed finally after integrating out
the $A_\mu^a$ field.  However, it generates the
additional complicated logarithmic determinant,
$
 \ln \det [D Q^{-1}D] .
$
Such a case was treated in \cite{QR97}.  
The choice of the gauge-fixing parameter should not change the
physics, since it appears due to a gauge choice.
Therefore we don't treat this case in this paper.
 }
 the Lagrange multiplier field
$\phi^a$ can be easily integrated out. 
The result is
\begin{eqnarray}
  \phi^a F^a[A,a] + {\alpha \over 2} (\phi^a)^2
+ \phi^a J_\phi^a 
  \rightarrow - {1 \over 2\alpha} (F^a[A,a])^2
- {1 \over \alpha} F^a[A,a] J_\phi^a .
\end{eqnarray}

\par
Next, as a preliminary procedure to integrate out
$A_\mu^a$, we rewrite the last term in the action
(\ref{apYM}) into
\begin{eqnarray}
   ({\cal S}_{\mu\nu}^a)^2 
   &=& - 2 A_\mu^a W_{\mu\nu}^{ab} A_\nu^b
   + 2 \partial_\mu(A_\nu^a {\cal S}_{\mu\nu}^a),
\nonumber\\
   W_{\mu\nu}^{ab} 
   &:=& (D^\rho[a] D_\rho[a])^{ab} \delta_{\mu\nu}
   - \epsilon^{ab3} f_{\mu\nu}
   - D_\mu[a]^{ac} D_\nu[a]^{cb} ,
\end{eqnarray}
where we have used 
\begin{eqnarray}
 [D_\mu[a]^{ac}, D_\nu[a]^{cb}] = - \epsilon^{ab3}
f_{\mu\nu} .
\label{fs}
\end{eqnarray}
Discarding the surface term, 
\footnote{
This will be justified, since the off-diagonal gluons
become massive  due to MAG.
}
we arrive at 
\begin{eqnarray}
   S_{YM} &=& S_{YM}[a,A,B,c,\bar c; J] =
   S_1[a,B] + S_2[a,c,\bar c] + S_3[a,A,B,c,\bar c;J] ,
    \label{apBFYM3}
\\
   S_1 &=&  \int d^4x \left[
    - {1 \over 4g^2}  f_{\mu\nu} f^{\mu\nu}
    - {1 \over 4} g^2 B_{\mu\nu} B^{\mu\nu}
\right]  ,
\label{S1}
\\
 S_2 &=&  \int d^4x \left[
 i \bar c^a D^\mu{}^{ac}[a]^\xi D_\mu^{cb}[a] c^b
+ i \bar c^3 \partial^\mu \partial_\mu c^3
+  \phi^3 (\partial^\mu a_\mu) + {\beta \over 2} (\phi^3)^2
\right]  ,
\label{S2}
\\
 S_3
&=&  \int d^4x \left[
  {1 \over 2g^2} A_\mu^a  Q_{\mu\nu}^{ab} A_\nu^b  
+ A_\mu^a  \left( G_\mu^a + 
{1 \over \alpha}  D^\mu{}^{ab}[a]^\xi J_\phi^b
 +  J^\mu{}^a \right) \right] ,
\label{S3}
   \\
     Q_{\mu\nu}^{ab} &:=&  
   (D_\rho[a] D_\rho[a])^{ab} \delta_{\mu\nu}
   - 2 \epsilon^{ab3} f_{\mu\nu}
 + {1 \over 2} g^2 \epsilon^{ab3}
   \epsilon_{\mu\nu\rho\sigma} B^{\rho\sigma} 
\nonumber\\&&
 - 2i g^2 \xi (\bar c^a  c^b - \bar c^c c^c \delta^{ab})
\delta_{\mu\nu}
 - D_\mu[a]^{ac} D_\nu[a]^{cb} 
 + {1 \over \alpha}  D_\mu[a]_\xi^{ac} D_\nu[a]_\xi^{cb} ,
\label{defK}
\\
 G_\mu^c  &:=& i (\partial_\mu \bar c^3 )
\epsilon^{cb3} c^b + i \bar c^a \epsilon^{ab3} [(1-\xi)
(\partial_\mu c^3)
\delta^{bc} - \xi \epsilon^{bc3} a_\mu c^3]
- i \partial_\mu(\bar c^a \epsilon^{ac3} c^3) .
    \label{defG}
\end{eqnarray}
where we have rescaled the parameter $\alpha$ to absorb the
$g$ dependence.
\par
All the terms appearing in the resulting YM action are  at
most  quadratic in $A_\mu^a$. 
Therefore the field $A_\mu^a (a=1,2)$ in $S_3$ can be
eliminated using the Gaussian integration and we obtain
\begin{eqnarray}
   i S_0
&=&  \ln \int [dA_\mu^a] \exp \left\{ i \int d^4x \left[
  {1 \over 2g^2} A_\mu^a  Q_{\mu\nu}^{ab} A_\nu^b  
+ A_\mu^a  \left( G_\mu^a + 
{1 \over \alpha}  D^\mu{}^{ab}[a]^\xi J_\phi^b
 +  J^\mu{}^a \right)
 \right] \right\}
\nonumber\\
&=& - {1 \over 2} \ln \det (Q_{\mu\nu}^{ab}) 
 + {g^2 \over 2}  G_\mu^a (Q^{-1})_{\mu\nu}^{ab} G_\nu^b
 + g^2 \left({1 \over \alpha}  D^\mu{}^{ac}[a]^\xi J_\phi^c
 +  J^\mu{}^a \right)  (Q^{-1})_{\mu\nu}^{ab} G_\nu^b 
\nonumber\\&&
 - {g^2 \over 2\alpha} J_\phi^a D^{ab}[a]^\xi 
(Q^{-1})_{\mu\nu}^{ab} D^{\nu}{}^{cd}[a]^\xi J_\phi^d
+ {g^2 \over \alpha} J^\mu{}^a (Q^{-1})_{\mu\nu}^{ab}
D^{\nu}{}^{bc}[a]^\xi J_\phi^c 
\nonumber\\&&
+ {g^2 \over 2} J^\mu{}^a (Q^{-1})_{\mu\nu}^{ab} J^\nu{}^b .
\label{S0}
\end{eqnarray}
Thus we obtain the effective abelian gauge theory
\begin{eqnarray}
  S_E &=& S_0[a,B,c,\bar c;J] + S_1[a,B] + S_2[a,c,\bar c],
\nonumber\\
 S_0 &=& - {1 \over 2} \ln \det (Q_{\mu\nu}^{ab})
 + {g^2 \over 2}  G_\mu^a (Q^{-1})_{\mu\nu}^{ab} G_\nu^b
 + g^2 \left({1 \over \alpha}  D^\mu{}^{ac}[a]^\xi J_\phi^c
 +  J^\mu{}^a \right)  (Q^{-1})_{\mu\nu}^{ab} G_\nu^b 
\nonumber\\&&
 - {g^2 \over 2\alpha} J_\phi^a D^{ab}[a]^\xi 
(Q^{-1})_{\mu\nu}^{ab} D^{\nu}{}^{cd}[a]^\xi J_\phi^d
+ {g^2 \over \alpha} J^\mu{}^a (Q^{-1})_{\mu\nu}^{ab}
D^{\nu}{}^{bc}[a]^\xi J_\phi^c 
\nonumber\\&&
+ {g^2 \over 2} J^\mu{}^a (Q^{-1})_{\mu\nu}^{ab} J^\nu{}^b .
   \label{effabelYM}
\end{eqnarray}
  As will be shown in the next subsection,
$\ln \det Q$ gives the renormalization of the fields
$a_\mu, B_{\mu\nu}$ and
$c^a$.  
The residual U(1) invariant theory is obtained by putting 
$\phi^3 = 0$ and $\bar c^3 = c^3 = 0$ (hence
$G_\mu^a = 0$).  Therefore, the resulting APEGT is greatly
simplified.
\par
On the other hand, the effective abelian BF-YM theory is
obtained if $S_1$ and 
$Q_{\mu\nu}^{ab}$ in $S_3$ are
replaced by
\begin{eqnarray}
   S_1 &=&  \int d^4x \left[
  {1 \over 4} \epsilon^{\mu\nu\rho\sigma}
B_{\rho\sigma}      f_{\mu\nu}  
    - {1 \over 4} g^2 B_{\mu\nu} B^{\mu\nu}
\right] ,
\nonumber\\
     Q_{\mu\nu}^{ab} &:=&  
   (D_\rho[a] D_\rho[a])^{ab} \delta_{\mu\nu}
   - \epsilon^{ab3} f_{\mu\nu}
 + {1 \over 2} g^2 \epsilon^{ab3}
   \epsilon_{\mu\nu\rho\sigma} B^{\rho\sigma} 
\nonumber\\&&
 - 2i g^2 \xi (\bar c^a  c^b - \bar c^c c^c \delta^{ab})
\delta_{\mu\nu}
 - D_\mu[a]^{ac} D_\nu[a]^{cb} 
 + {1 \over \alpha}  D_\mu[a]_\xi^{ac} D_\nu[a]_\xi^{cb} ,
   \label{effapBFYM}
\end{eqnarray}
where the $G$ is the same as (\ref{defG}).
This case is discussed in Appendix A.

\subsection{Calculation of logarithmic determinant}

In MAG ($\xi=1$), the last two terms in $Q$ cancel by
taking
$\alpha=1$ (they disappear also for $\alpha=0$
\cite{QR97}), 
\begin{eqnarray}
     Q_{\mu\nu}^{ab} &:=&  
   (D_\rho[a] D_\rho[a])^{ab} \delta_{\mu\nu}
   - 2 \epsilon^{ab3} f_{\mu\nu}
 + {1 \over 2} g^2 \epsilon^{ab3}
   \epsilon_{\mu\nu\rho\sigma} B^{\rho\sigma} 
\nonumber\\&&
 - 2i g^2 (\bar c^a  c^b - \bar c^c c^c \delta^{ab})
\delta_{\mu\nu} .
 \label{K0}
\end{eqnarray}
\par
In order to calculate the $\ln \det Q$, we use the $\zeta$
function regularization or heat kernel method
(see e.g. \cite{DPY84}),
\begin{eqnarray}
  \ln \det Q = - \lim_{s \rightarrow 0} {d \over ds}
{\mu^{2s} \over \Gamma(s)} \int_{0}^{\infty} dt \ t^{s-1}
\Tr(e^{-t Q}) ,
\label{hk}
\end{eqnarray}
where $\Tr$ is understood in the functional sense.
In this subsection the calculations are performed in
Euclidean formulation.
\par
First, we calculate the trace of $e^{-t Q}$.
To estimate this quantity, we use the plane wave basis,
\begin{eqnarray}
  \Tr(e^{-t Q}) = \int d^4x \ \tr \langle x| e^{-t Q} |x
\rangle  
= \int d^4x \ \tr \int {d^4k \over (2\pi)^4} e^{-ikx}
e^{-t Q}  e^{ikx} .
\label{hk1}
\end{eqnarray}
By making use of the relation,
\begin{eqnarray}
  [D_\mu^{ab},  e^{\pm ikx} ] = \pm i k_\mu e^{\pm ikx}
\delta^{ab},
\end{eqnarray}
we find
\begin{eqnarray}
  e^{-ikx} e^{-t (D_\rho[a]^2)^{ab} \delta_{\mu\nu}} 
e^{ikx} 
  = \exp [-t
  (D_\rho[a]^{ac} + ik_\rho \delta^{ac})(
  D_\rho[a]^{cb} + ik_\rho \delta^{cb}) \delta_{\mu\nu}] .
\end{eqnarray}
Furthermore the rescaling of $k_\mu$, 
$k_\mu \rightarrow k_\mu/\sqrt{t}$, leads to
\begin{eqnarray}
  \Tr(e^{-t Q})  
&=&  \int d^4x {1 \over t^{2}} \tr \int {d^4k \over
(2\pi)^4} e^{k_\mu k^\mu} \exp [-(2i \sqrt{t} k^\mu D_\mu +
t Q)]
\nonumber\\
&=& \int d^4x {1 \over t^{2}} 
\sum_{n=0}^{\infty} {(-1)^n \over n!}
\tr \int {d^4k \over (2\pi)^4} e^{k_\mu k^\mu}
 (2i \sqrt{t} k^\mu D_\mu + t Q)^n  ,
\end{eqnarray}
where we have omitted the unit operator,
$\delta_{ab}\delta_{\mu\nu}$.
It is obvious that all terms odd w.r.t. $k_\mu$ in the
expansion go to zero in the  integration.
Thus we obtain
\footnote{
The zero-order term of the expansion with respect to
$t$ is equal to the free term 
\begin{eqnarray}
 \Tr(\exp[ -t Q_0] :=
 \Tr(\exp[-t \partial^2 \delta^{ab} \delta_{\mu\nu}])
= {4 N(N-1) \int d^4x \over 16\pi^2 t^2} .
\end{eqnarray}
}
\begin{eqnarray}
 && \Tr(e^{-t Q}) - \Tr(e^{-t Q_0}) 
\nonumber\\
&=& \int {d^4 x \over 16\pi^2}  \ \tr \left[{1 \over 2} Q^2
- D^2 Q + {1 \over 6} (2 D^2 D^2 + D_\mu D_\nu D_\mu D_\nu)
\right] + O(t) ,
\end{eqnarray}
where we have used the cyclicity of trace and the
replacement
\begin{eqnarray}
   k_\mu k_\nu &\rightarrow& 
{1 \over 4} k^2 \delta_{\mu\nu} ,
\nonumber\\
 k_\mu k_\nu k_\alpha k_\beta &\rightarrow& {1 \over 24}
(k^2)^2 (g_{\mu\nu}g_{\alpha\beta}+g_{\mu\alpha}g_{\nu\beta}
+g_{\mu\beta}g_{\nu\alpha}) ,
\end{eqnarray}
which is applied in the integrand of the integration formula
\begin{eqnarray}
   \int {d^4k \over (2\pi)^4} e^{k^2} (k^2)^m 
= {(-1)^m \over 16\pi^2} (m+1)! \ (m=0, 1, 2, \cdots) .
\end{eqnarray}
Separating the first term from the other terms in $Q$,
\begin{eqnarray}
     Q_{\mu\nu}^{ab} &:=&  
   (D_\rho[a] D_\rho[a])^{ab} \delta_{\mu\nu}
   + \tilde  Q_{\mu\nu}^{ab},
 \label{K01}
\end{eqnarray}
we see
\begin{eqnarray}
 && \Tr(e^{-t Q}) - \Tr(e^{-t Q_0}) 
\nonumber\\
&=& {1 \over 16\pi^2} \int d^4 x \ \tr[{1 \over 2}\tilde
Q^2 + {1 \over 6} D_\mu D_\nu (D_\mu D_\nu - D_\nu D_\mu)]
+ O(t) 
\nonumber\\
&=& {1 \over 16\pi^2}  \int d^4 x \ \tr \left( {1 \over
2}\tilde Q^2 + {1 \over 12} [D_\mu,D_\nu] [D_\mu, D_\nu]
\right) + O(t) ,
\end{eqnarray}
where any cross term between $D$ and $\tilde Q$ does not
appear. 
\par
The first term is obtained as
\begin{eqnarray}
 \tr \left( {1 \over 2}\tilde Q^2 \right)
&=&  2  \kappa f_{\mu\nu}f^{\mu\nu}
-  {1 \over 2} g^4 \kappa B_{\mu\nu}B^{\mu\nu}
-  \kappa  g^2 \epsilon^{\mu\nu\rho\sigma} 
B_{\rho\sigma} f_{\mu\nu} 
\nonumber\\&&
- 8 g^4 (\bar c^a  c^b - \bar c^c c^c \delta^{ab})
(\bar c^b  c^a - \bar c^d c^d \delta^{ba}),
\end{eqnarray}
and the second term is
\begin{eqnarray}
 \tr \left({1 \over 12} [D_\mu,D_\nu][D_\mu, D_\nu] \right)
  = {-1 \over 3} \kappa f_{\mu\nu}f^{\mu\nu} ,
\end{eqnarray}
where
\begin{eqnarray}
    \kappa := C_2(G) := f^{3cd}f^{3cd} = 2 .
\end{eqnarray}
Thus we obtain (apart from the 4-body ghost
interaction terms, see Appendix B) the U(1) invariant
result,
\begin{eqnarray}
 {1 \over 2} \ln \det  Q_{\mu\nu}^{ab} 
  =   \int d^4x \left[
  {1 \over 4g^2} z_a f_{\mu\nu} f^{\mu\nu}
    + {1 \over 4} z_b g^2 B_{\mu\nu} B^{\mu\nu}
    +  {1 \over 2} z_c  
B_{\mu\nu} \tilde f_{\mu\nu} + \cdots \right] ,
\end{eqnarray}
where 
\begin{eqnarray}
 z_a = - {20 \over 3} \kappa {g^2 \over 16\pi^2} \ln \mu , 
\quad
 z_b = + 2 \kappa {g^2 \over 16\pi^2} \ln \mu ,
\quad
 z_c = + 4 \kappa {g^2 \over 16\pi^2} \ln \mu .
 \label{z}
\end{eqnarray}

\par
Therefore, in the absence of the source $J_\mu^a = 0 =
J_\phi^a$, 
\begin{eqnarray}
  S_0 + S_1
=
  \int d^4x \left[
    - {1+z_a \over 4g^2} f_{\mu\nu} f^{\mu\nu}
    - {1+z_b \over 4} g^2 B_{\mu\nu} B^{\mu\nu}
    +  {1 \over 2} z_c  
B_{\mu\nu} \tilde f_{\mu\nu} + \cdots 
   \right]  .
\end{eqnarray}
Integrating out the $B_{\mu\nu}$ field, we will obtain an
additional contribution,
\begin{eqnarray}
 - {1 \over 4g^2} z_c^2 (1+z_b)^{-1} f_{\mu\nu} f^{\mu\nu}
.
\end{eqnarray}
However, in the one-loop level, this term is irrelevant.
Therefore, the cross term does not contribute in the
one-loop level.

\par
For later convenience, we calculate another determinant
coming from the integration over ghost fields.
For the action,
\begin{eqnarray}
 S_F = \int d^4x \
i \bar c^a D_\mu^{ac}[a] D_\mu^{cb}[a] c^b ,
\end{eqnarray}
we obtain up to one-loop
\begin{eqnarray}
S_c &=&  \ln \int [d\bar c][dc] \exp \left\{ - \int d^4x \
\bar c^a D_\mu^{ac}[a] D_\mu^{cb}[a] c^b
\right\}
\nonumber\\
&=& \ln \det (D_\mu^{ac}[a] D_\mu^{cb}[a]) 
\nonumber\\
&=&  \int d^4x {1 \over 4g^2} z_a' f_{\mu\nu} f^{\mu\nu}
+ \cdots ,
\quad
z_a' :=  {2 \over 3} \kappa {g^2 \over 16\pi^2} \ln \mu .
\label{lndetc}
\end{eqnarray}
\par

For the abelian-projected effective BF-YM theory, see
Appendix A.

\subsection{APEGT with monopole}
\par
 The antisymmetric (abelian) tensor
$B_{\mu\nu}$ has the  Hodge decomposition in 3+1
dimensions (see section 6 for other dimensions),
\begin{eqnarray}
  B_{\mu\nu} = b_{\mu\nu} + \tilde \chi_{\mu\nu} , 
  \quad
 b_{\mu\nu} := \partial_\mu b_\nu - \partial_\nu b_\mu .
  \quad
  \tilde \chi_{\mu\nu}= {1 \over 2}
\epsilon_{\mu\nu\alpha\beta}
 (\partial^\alpha \chi^\beta - \partial^\alpha \chi^\beta) .
 \label{Hd}
\end{eqnarray}
The tensor $B_{\mu\nu}$ has six degrees of freedom, while
the fields $b_\mu$ and $\chi_\mu$ have eight.  This mismatch
is not a problem, since two degrees are redundant; the
gauge transformation 
\begin{eqnarray}
  b_\mu(x) \rightarrow b_\mu'(x) = b_\mu(x) - \partial_\mu
\theta,
\quad
  \chi_\mu(x) \rightarrow \chi_\mu'(x) = \chi_\mu(x) -
\partial_\mu
\varphi,
\label{dualGT}
\end{eqnarray}
leave $B_{\mu\nu}$ invariant.
In the function integral, the integration over $B_{\mu\nu}$
is replaced by an integration over $b_\mu$ and $\chi_\mu$,
provided that the gauge degrees of freedom are fixed in
(\ref{dualGT}).  These gauge fixing are not explicitly
presented in the following, since they can be easily
implemented.
\par
In this case, we obtain
\begin{eqnarray}
  S_0 + S_1
&=&
  \int d^4x \Big[
    - {1+z_a \over 4g^2}  f_{\mu\nu} f^{\mu\nu}
    - {1+z_b \over 4} g^2 (b_{\mu\nu} b^{\mu\nu}
 + \tilde \chi_{\mu\nu} \tilde \chi^{\mu\nu})
 \nonumber\\
 &&+ {1 \over 2} z_c b_{\mu\nu} \tilde f_{\mu\nu} 
    +  {1 \over 2} z_c \chi_{\mu\nu} f_{\mu\nu}  
     + \cdots 
   \Big]  ,
\end{eqnarray}
At one-loop level, integration over $\chi$ leads to
\begin{eqnarray}
  S_E
=
  \int d^4x \left[
    - {1+z_a \over 4g^2}  f_{\mu\nu} f^{\mu\nu}
 + i \bar c^a D_\mu^{ac}[a] D_\mu^{cb}[a] c^b 
 - {1+z_b \over 4} g^2 b_{\mu\nu} b^{\mu\nu}
 - z_c b_\mu k^\mu 
 \right]  ,
 \label{APEGT}
\end{eqnarray}
where we have defined the magnetic current,
\begin{eqnarray}
  k^\mu := \partial^\nu \tilde f_{\mu\nu},
\quad \tilde f_{\mu\nu} :=  {1 \over 2} 
\epsilon_{\mu\nu\rho\sigma}f^{\rho\sigma} .
\end{eqnarray}
Here we have neglected the ghost self-interaction
terms (see Appendix B) and higher derivative terms coming
from the logarithmic determinant of $Q$. This is
the APEGT written in terms of the abelian gauge field
$a_\mu$ and the dual gauge field $b_\mu$ (an effect of
the off-diagonal ghost field is studied in the next
section).  This theory has
$U(1)_e \times U(1)_m$ symmetry where
 the abelian gauge field $a_\mu$ has $U(1)_e$ symmetry and
the dual abelian gauge field $b_\mu$ has $U(1)_m$ symmetry
which is guaranteed by the conservation 
$
 \partial_\mu k^\mu = 0 .
$
If the field
$a_\mu$ is singular, the magnetic current $k_\mu$ is
non-zero and couples with the dual field $b_\mu$.  
This interaction leads
to the dual Meissner effect, see section 4.
In the absence of magnetic current, the dual field $b_\mu$
decouples from the theory.
Note that the renormalizations of the fields $a_\mu, b_\mu$
are different each other.
\par
APEGT can be considered as an interpolating theory which
reduces to a theory with an action $S[a]$ by integrating
out the
$b_\mu$ field or to another theory with  $S[b]$ by
integrating out $a_\mu$ field.
The theory $S[a]$ is suitable for describing the weak
coupling region, while $S[b]$ is more suitable for the
strong coupling region.
However, both theories give the dual description of the
same physics.  
In the next section, we see an aspect of this picture.

\section{One-loop beta function and asymptotic freedom}
\setcounter{equation}{0}

Neglecting  the contribution from dual gauge field, the
APEGT is reduced to the $U(1)$ gauge theory,  
\begin{eqnarray}
  S_E  =
  \int d^4x \left[
    - {1+z_a \over 4g^2}  f_{\mu\nu} f^{\mu\nu}{}
+ i \bar c^a D_\mu^{ac}[a] D_\mu^{cb}[a] c^b \right] .
\label{APEGT1}
\end{eqnarray}
This APEGT is similar to the scalar quantum electrodynamics.
But the scalar field is replaced with the ghost field.
We can show that the running coupling $g$ exhibits asymptotic
freedom, i.e. the beta function has negative
coefficient. The beta function is obtained from the
calculation of the logarithmic determinant in the previous
section.
\par
We define the wave function renormalization for 
$a_\mu$ and $c^a$ by
\begin{eqnarray}
  a_\mu^R = Z_a^{-1/2} a_\mu ,
\quad
  c_R = Z_c^{-1/2} c  .
\end{eqnarray}
\par
For the 3-point $a_\mu c \bar c$ vertex, the renormalized
coupling constant is defined by 
\begin{eqnarray}
 g_R = Z_a^{1/2} Z_c Z_g^{-1} g  .
\end{eqnarray}
It should be remarked that the effective abelian gauge
theory (\ref{APEGT1}) has U(1) gauge invariance and we can
derive the Ward-Takahashi (WT) identity for this symmetry. 
For example, the 3-point vertex function and the ghost
propagator obeys the well known WT identity which is
similar to that in scalar QED. This implies that
$Z_g=Z_c$ (independently on the order of the perturbation).
Therefore the coupling constant for the
$a_\mu c
\bar c$ vertex is determined by $Z_a$ alone,
\begin{eqnarray}
 g_R = Z_a^{1/2} g  .
\end{eqnarray}
Note that $Z_a$ is obtained by integrating the ghost field,
i.e. $\ln \det D^2$, if we remember (\ref{lndetc}) .
Adding this contribution to (\ref{APEGT1}), we obtain
\begin{eqnarray}
 Z_a = 1 - z_a + z_a'
 = 1 + {g^2 \over 16\pi^2} {22C_2(G) \over 3} \ln \mu,
 \quad
  C_2(G)   := f^{3cd}f^{3cd} = 2 .
\end{eqnarray}
Thus the $\beta$-function is easily calculated:
\begin{eqnarray}
  \beta(g) := \mu {dg_R \over d\mu} 
= - {b_0 \over 16\pi^2} g_R^3, 
\quad  b_0 = {11C_2(G) \over 3} > 0 .
\end{eqnarray}
Thus the APEGT exhibits asymptotic
freedom as the original YM theory.
\footnote{
This fact was first obtained in the
gauge $\alpha=0$ based on quite complicated calculations 
\cite{QR97}.
}
\par
In order to obtain the RG beta function, we could have used
the Feynman graph technique.  
By the perturbation expansion in the coupling constant, we
can ascertain the Ward relation $Z_g=Z_c$.
\footnote{
Explicit calculation based on the perturbation theory
shows that
\begin{eqnarray}
 Z_g = Z_c = 1 - {g^2 \over 16\pi^2} 2(\beta-3) \ln \mu,
\end{eqnarray}
where $\beta$ is the gauge-fixing parameter.
}
An origin of asymptotic freedom ($z_a'$) is understood as
follows.  By the Ward relation, asymptotic freedom is
explained by the vacuum polarization of the abelian gauge
field alone.  This diagram up to order $g^2$ is quite
similar to those of scalar QED by replacing the complex
scalar field $\phi,
\phi^*$ with the ghost, antighost field $c^a, \bar c^a$;
\begin{eqnarray}
 \bar c^a D_\mu^{ab}[a] D_\mu^{bc}[a] c^c 
\leftrightarrow |(\partial_\mu - ie a_\mu) \phi|^2 
= - \phi^* (\partial_\mu - ie a_\mu)^2 \phi.
   \label{corresp}
\end{eqnarray}
An essential difference is the
signature due to ghost loop.  This minus sign changes the
non asymptotic freedom of scalar QED into asymptotic
freedom in the effective abelian gauge theory in
question.  The additional dominant contribution
($z_a$) comes from the gluon self-interaction which is
already included in the action of APEGT through the
calculation of 
$-(1/2) \ln \det Q$.  Summation of two contributions gives
exactly the same beta function as the original YM theory.
\par
In other words, the APEGT is the abelian gauge theory with
QCD-like running coupling constant $g(\mu)$,
\begin{eqnarray}
  S_E[a]
  =  \int d^4x \left[
    - {1 \over 4g(\mu)^2}  f_{\mu\nu}  f^{\mu\nu} 
  \right] ,
  \quad
  {1 \over g(\mu)^2} =  {1 \over g(\mu_0)^2}
  + {b_0 \over 8\pi^2} \ln {\mu \over \mu_0} .
  \label{action1}
\end{eqnarray}

\section{Monopole condensation and dual Meissner effect}
\setcounter{equation}{0}

In section 2, we have obtained the APEGT with magnetic
current (after the ghost integration),
\begin{eqnarray}
  S_E[a,b,k]
  =  \int d^4x \left[
    - {1 \over 4g^2}  f_{\mu\nu}^R f^R{}^{\mu\nu} 
    - {1 \over 4} b_{\mu\nu}^R b^R{}^{\mu\nu} 
 - {1 \over g} z_c/Z_b^{1/2} b_\mu^R k^\mu  \right] .
\label{APEGT2}
\end{eqnarray}
The interaction term between the dual gauge field $b_\mu$ and
the magnetic current $k_\mu$ is generated by the
radiative correction through the gluon self-interaction.
The action leads to the field equation for the
renormalized field,
\begin{eqnarray}
  \partial_\mu f_R^{\mu\nu}  = j_{R}^\nu,
  \quad
  \partial_\mu b_R^{\mu\nu}  = k_{R}^\nu,
\end{eqnarray}
where we have defined
\begin{eqnarray}
 k_{R}^\mu := {1 \over g} (z_c/Z_b^{1/2})  k^\mu ,
 \quad
 Z_b^{1/2} = 1 - z_b/2 .
\end{eqnarray}
\par
Integrating out the dual field $b_\mu$, we obtain the
effective action for the monopole loop,
\begin{eqnarray}
  S_E[a,k]
\cong
  \int d^4x \left[
    - {Z_a^{-1} \over 4g^2}  f_{\mu\nu} f^{\mu\nu}
+ {1 \over g^2} k^\mu D_{\mu\nu} k^\nu 
\right]  ,
\label{maction}
\end{eqnarray}
where $D_{\mu\nu}$ is the massless vector propagator.
Such a monopole action was predicted on a lattice in
\cite{SS91}.
\par
For our purposes, it is more convenient to use the local
lagrangian formalism invented by Zwanziger
\cite{Zwanziger71} for the system having both electric and
magnetic currents.
\par
Before that, we will give a different treatment which is
helpful to discuss the relationship
between the monopole condensation and the instanton. 
We show how the magnetic monopole current is calculated in
the original YM theory.

\subsection{Definition of the monopole current}
\par
We show that the current $K_\mu$ defined
by 
\begin{eqnarray}
  K^\mu = {1 \over 2}  
  \epsilon^{\mu\nu\rho\sigma} \partial_\nu (
 \epsilon^{ab3} A_\rho^a A_\sigma^b)
= {1 \over 2}  
  \epsilon^{\mu\nu\rho\sigma} \partial_\nu  
 {\cal C}_{\rho\sigma}
\label{defmc}
\end{eqnarray}
is interpreted as the magnetic monopole current. 
This current is topologically conserved, i.e.,
$
\partial_\mu K^\mu = 0 .
$
For a while, we use a different normalization of the field
${\cal A} \rightarrow g{\cal A}$.
Usually the abelian gauge field $a_\mu$ defined by
$
 a_\mu(x) := \tr[T^3 {\cal A}_\mu(x)]
$
can have singularities if the field ${\cal A}$ is gauge
transformed by the rotation matrix $U(x)$
as
$
 {\cal A}_\mu(x) \rightarrow {\cal A}_\mu^U(x) ,
$
\begin{eqnarray}
{\cal A}_\mu^U(x) := 
 U(x) {\cal A}_\mu(x) U^\dagger(x) 
 + {i \over g}
U(x) \partial_\mu U^\dagger(x) ,
\label{GT}
\end{eqnarray}
such that the gauge
transformed field ${\cal A}_\mu^U(x)$ satisfies the
abelian gauge fixing condition, e.g. MAG.
It is this singularity that leads to a non-zero magnetic
current.  
Under the gauge transformation (\ref{GT}), the field
strength is transformed 
as
$
 {\cal F}_{\mu\nu}(x) \rightarrow {\cal F}_{\mu\nu}^U(x) ,
$
\begin{eqnarray}
 {\cal F}_{\mu\nu}^U(x) &=& 
 U(x) {\cal F}_{\mu\nu}(x) U^\dagger(x) 
 \nonumber\\
&=&   \partial_\mu {\cal A}_\nu^U(x) 
 -   \partial_\nu {\cal A}_\mu^U(x)
 - ig [{\cal A}_\mu^U(x), {\cal A}_\nu^U(x)] ,
\label{GT2}
\end{eqnarray}
see Appendix C.
The abelian gauge field strength is extracted as
\begin{eqnarray}
 f_{\mu\nu} 
 &:=& \partial_\mu a_\nu^U - \partial_\nu a_\mu^U 
= \tr[T^3 (\partial_\mu {\cal A}_\nu^U
- \partial_\nu {\cal A}_\mu^U)]
\label{afs}
\\
&=&  \tr\left[T^3 \left( U {\cal F}_{\mu\nu} U^\dagger
+ ig [{\cal A}_\mu^U, {\cal A}_\nu^U]) 
\right) \right] .
 \label{GT3}
\end{eqnarray}
The definition of the magnetic current is  
\begin{eqnarray}
 k_\mu := {1 \over 2} \epsilon_{\mu\nu\rho\sigma}
\partial^\nu f^{\rho\sigma} .
 \label{mc}
\end{eqnarray}
The first term in (\ref{GT}) is non-singular.  Hence
(\ref{afs}) shows that the first term gives vanishing
contribution in the magnetic current.  Only the second
term 
\begin{eqnarray}
  \tilde {\cal A}_\mu(x) :=  {i \over g}
U(x) \partial_\mu U^\dagger(x)
\end{eqnarray}
give a non-vanishing magnetic current. 
If $U(x)$ is not singular, $\tilde {\cal A}_\mu$ is a pure
gauge and hence the field strength constructed from
$\tilde {\cal A}_\mu$ is zero,
$
 \tilde {\cal F}_{\mu\nu}(x) :=
 \partial_\mu \tilde {\cal A}_\nu(x) 
 -   \partial_\nu \tilde {\cal A}_\mu(x)
 - i g [\tilde {\cal A}_\mu(x), \tilde {\cal A}_\nu(x)] 
 \equiv 0 .
$
For the singular $U(x)$, this is modified as
\begin{eqnarray}
 \tilde {\cal F}_{\mu\nu}(x) :=
 \partial_\mu \tilde {\cal A}_\nu(x) 
 -   \partial_\nu \tilde {\cal A}_\mu(x)
 - i g[\tilde {\cal A}_\mu(x), \tilde {\cal A}_\nu(x)]
 = {i \over g} U(x)[\partial_\mu, \partial_\nu]
U^\dagger(x)  .
\label{GT4}
\end{eqnarray}
Thus we obtain the expression of the magnetic current,
\begin{eqnarray}
 k_\mu &=& {1 \over 2}
 \epsilon_{\mu\nu\rho\sigma} \partial^\nu 
  \tr[T^3 (\partial_\rho \tilde {\cal A}_\sigma
- \partial_\sigma \tilde {\cal A}_\rho)]
\nonumber\\
&=& {1 \over 2} \epsilon_{\mu\nu\rho\sigma} \partial^\nu 
  \tr(T^3 ig[\tilde {\cal A}_\rho, \tilde {\cal
A}_\sigma] )   
+ {1 \over 2} \epsilon_{\mu\nu\rho\sigma} \partial^\nu 
  \tr(T^3 {i \over g} U[\partial_\rho, \partial_\sigma]
U^\dagger  ) .
 \label{mcf}
\end{eqnarray}
The magnetic current is composed of two parts.
The second part corresponds to the contribution from the
Dirac string.
Therefore the first part is the contribution from the
magnetic monopole which agrees with (\ref{defmc}) in the
original normalization of the field ${\cal A}$. This can be
seen also from (\ref{GT3}),
since
\begin{eqnarray}
 k_\mu 
 &=& {1 \over 2} \epsilon_{\mu\nu\rho\sigma} \partial^\nu 
  \tr(T^3 ig[\tilde {\cal A}_\rho, \tilde {\cal
A}_\sigma] )   
+ {1 \over 2} \epsilon_{\mu\nu\rho\sigma} \partial^\nu 
  \tr(T^3 {\cal F}_{\mu\nu}^U  )  
\nonumber\\
&=& {1 \over 2} \epsilon_{\mu\nu\rho\sigma} \partial^\nu 
  \tr(T^3 ig[\tilde {\cal A}_\rho, \tilde {\cal
A}_\sigma] )   
+ {1 \over 2} \epsilon_{\mu\nu\rho\sigma} \partial^\nu 
  \tr(T^3 {i \over g} U[\partial_\rho, \partial_\sigma]
U^\dagger  ) .
 \label{mcf2}
\end{eqnarray}
For details, see Appendix C.

\subsection{Dual effective abelian theory}
\par
In the following we present somewhat different picture of
monopole condensation leading to the dual Meissner effect.
By extracting the $b_\mu$ dependent pieces from the action 
(\ref{apBFYM3}), and  
inserting the identity
\begin{eqnarray}
  1  =  \int [dK^\mu] \delta(K^\mu - {1 \over 2}  
  \epsilon^{\mu\nu\rho\sigma} \partial_\nu (
 \epsilon^{ab3} A_\rho^a A_\sigma^b)) , 
\end{eqnarray}
the partition function $Z_{YM}$ is written as
\begin{eqnarray}
  Z_{YM}[J] &:=& \int d\mu e^{-S_{YM}}
  = \int d\mu \int [dK^\mu]
  \delta(K^\mu - {1 \over 2}   
  \epsilon^{\mu\nu\rho\sigma} \partial_\nu (
 \epsilon^{ab3} A_\rho^a A_\sigma^b)) 
 \nonumber\\&&
  \times \exp \left\{ -S_{YM}[a,A,\chi,c,\bar c;J] -
  \int d^4x \left[ - {1 \over 4} g^2 b_{\mu\nu} b^{\mu\nu}
  + b^\mu K_\mu  \right] \right\} ,
\end{eqnarray}
where the measure $d\mu$ denotes the integration over 
all the fields.
\par
In order to see that the APEGT can exhibit dual Meissner
effect, we consider the effective action $S[b]$ written in
terms of
$b_\mu$ which is obtained by integrating out all the fields
except for $b_\mu$,
\begin{eqnarray}
  Z_{YM}[J] &:=& \int [db_\mu] 
  \exp \left\{ - S[b]  \right\} .
\end{eqnarray}
Then $S[b]$ is obtained as
\begin{eqnarray}
   S[b] = {-1 \over 4}g^2 \int d^4x b_{\mu\nu} b^{\mu\nu}
+   \ln \langle \exp [\int d^4x b_\mu(x) K_\mu(x) ]
\rangle_0 ,
\end{eqnarray}
where the expectation value for a function $f$ of the field is
defined by
\begin{eqnarray}
 \langle  f(A)  \rangle_0 &:=& 
\int d\tilde \mu \int [dK^\mu]
  \delta(K^\mu - {1 \over 2}  
  \epsilon^{\mu\nu\rho\sigma} \partial_\nu (
 \epsilon^{ab3} A_\rho^a A_\sigma^b)) 
 \nonumber\\&&
  \times \exp \left\{ -S_{YM}[a,A,\chi,c,\bar c;J] \right\} 
f(A),
\end{eqnarray}
where $d\tilde \mu$ denotes the normalized measure without
$[db_\mu]$ so that
$
 \langle 1 \rangle_0 = 1.
$
It turns out that 
\begin{eqnarray}
   S[b] &=& {-1 \over 4}  g^2
   \int d^4x b_{\mu\nu}(x) b^{\mu\nu}(x)
+  \int d^4x \langle K_\mu(x) \rangle_0  b^\mu(x)
\nonumber\\&&
+ {1 \over 2} \int d^4x \int d^4y
\langle K_\mu(x) K_\nu(y) \rangle_c b^\mu(x) b^\nu(y) 
+ O(b^3) ,
\label{dualaction}
\end{eqnarray}
where $\langle K_\mu(x) K_\nu(y) \rangle_c$ is the
connected correlation function obtained from the normalized
expectation value, 
$
 \langle f(A) \rangle :=
 \langle f(A) \rangle_0/\langle 1 \rangle_0 ,
$
e.g.,
$
\langle f(A) g(A) \rangle_c
= \langle f(A) g(A) \rangle
- \langle f(A) \rangle \langle g(A) \rangle .
$
\par
We can obtain a similar expression for the APEGT
using the action (\ref{APEGT}).  Hence the argument in
the next subsection can be extended also to the APEGT.

\subsection{Dual Meissner effect due to monopole condensation}

The effective dual abelian theory $S[b]$ has $U(1)$
symmetry, 
$
 b_\mu \rightarrow b_\mu + \partial_\mu \theta ,
$
which is different from the $U(1)$ symmetry for the
abelian field $a_\mu$ and is called the magnetic
$U(1)_m$ symmetry hereafter. The magnetic current
 satisfies the conservation
$
 \partial_\mu K^\mu = 0 ,
$
consistently with the $U(1)_m$ symmetry.  This implies that
the correlation function of the magnetic monopole current
is transverse,
\begin{eqnarray}
  \langle K_\mu(x) K_\nu(y) \rangle_c 
=  \left( \delta_{\mu\nu} \partial^2 
- {\partial_\mu \partial_\nu} \right)  M(x-y)    .
\label{kk}
\end{eqnarray}
As long as the magnetic $U(1)_m$ symmetry is not broken, 
the dual gauge field $b_\mu$ is always massless as can
be seen from (\ref{dualaction}) and (\ref{kk}).   Therefore
non-zero mass for the dual gauge field implies  breakdown
of the
$U(1)_m$ symmetry.
\par
If $U(1)_m$ symmetry is broken in such a way that
\begin{eqnarray}
  \langle K_\mu(x) K_\nu(y) \rangle_c 
=   g^2 \delta_{\mu\nu} \delta^{(4)}(x-y) f(x)
+ \cdots ,
\end{eqnarray}
 the mass term is generated,
\begin{eqnarray}
  S[b] = \int d^4x \left[ 
 {-1 \over 4} g^2  b_{\mu\nu}(x) b^{\mu\nu}(x)
  +  {1 \over 2} g^2 m_b^2 b_\mu(x) b_\mu(x) 
  + \cdots \right] ,
  \label{actionb}
\end{eqnarray}
if we write $f(x) = m_b^2$.
This can be called the dual Meissner effect;
the dual gauge field acquires a mass given
by
\begin{eqnarray}
  m_b^2 = {1 \over 4g^2} \Phi(0) ,
\end{eqnarray}
if the monopole  loop  condensation
occurs in the sense that,
\begin{eqnarray}
 \Phi(x) := \lim_{y \rightarrow x} 
  {\langle K_\mu(x) K_\mu(y) \rangle_c \over 
  \delta^{(4)}(x-y)}
\not= 0 .
\label{op}
\end{eqnarray}
This is a criterion of dual superconductivity of QCD.
\footnote{
For other proposals, see \cite{DSuper} and references
therein. }
It is consistent with the picture of dual superconductor
scenario for quark confinement proposed by Nambu, 't Hooft
and Mandelstam
\cite{Nambu74,tHooft81,Mandelstam76}. 
In the translation invariant theory, $\Phi(x)$
is an $x$-independent constant which depends only on the
gauge coupling constant $g$.
If we take a specific classical configuration to estimate
them, $x$-dependence may appear, see the effective dual
GL theory in the latter half of this subsection.
\par
It should be remarked that
$\Phi$ is not the local order parameter in the usual
sense.  In order to find the non-zero value of
$m_b$, we must extract,
from the magnetic monopole current correlation
function
$
\langle K_\mu(x) K_\nu(y) \rangle_c ,
$
a piece which is proportional to
the Dirac delta function $\delta^{(4)}(x-y)$ diverging as
$y
\rightarrow x$.
Therefore, if such type of strong short-range correlation
between two magnetic monopole loops does not exist, $\Phi$
is obviously zero.  This observation seems to be consistent
with the result of lattice simulations.
The monopole loops exist both in the confinement and the
deconfinement phases.
However, in the deconfinement phase
the monopole currents are dilute and the vacuum contains
only short monopole loops with some non-zero density.
In the confinement phase, on the other hand, the monopole
trajectories form the infinite long loops and the monopole
currents form a dense cluster, although there is a number
of small mutually disjoint clusters \cite{SS94}. 
\par
It should be remarked that APEGT doesn't need any scalar
field.  In this sense, the mechanism in which the dual
gauge field acquires a mass is different from the dual
Higgs mechanism.  Nevertheless, we can always introduce the
scalar field into APEGT so as to recover the spontaneously
broken
$U(1)_m$ symmetry,
\begin{eqnarray}
   {1 \over 2} m_b^2 b_\mu(x) b_\mu(x) 
   \rightarrow {1 \over 2} m_b^2 
   (b_\mu(x) -  \partial_\mu \theta (x))^2
   =   | (\partial_\mu - ib_\mu(x)) \phi(x)|^2 ,
\end{eqnarray}
where we identify 
\begin{eqnarray}
   \phi(x) = {m_b \over \sqrt{2}} e^{i\theta(x)} .
   \label{rfs}
\end{eqnarray}
Indeed, the  result is invariant under 
$ b_\mu \rightarrow b_\mu + \partial_\mu \alpha$ and
$ \theta \rightarrow \theta + \alpha$
($ \phi \rightarrow e^{i\alpha} \phi $).
Such a scalar field is called the
St\"uckelberg field or Batalin-Fradkin field
\cite{Kondo95}.
The case (\ref{rfs}) is obtained as an extreme type
II limit (London limit),
\begin{eqnarray}
   \lim_{\lambda \rightarrow \infty} V(\phi), \quad
    V(\phi) := \lambda (|\phi(x)|^2 - m_b^2/2)^2 ,
\end{eqnarray}
or non-linear $\sigma$ model with a constraint,
$
 \delta(|\phi(x)|^2 - m_b^2/2) .
$
The value $\phi_0$ at which the potential $V(\phi)$ has a
minimum is proportional to the mass $m_b$ of dual gauge
field,
\begin{eqnarray}
  m_b = \sqrt{2} \phi_0 = {\sqrt{\Phi} \over 2g} .
\end{eqnarray}
In the deconfinement phase, the minimum is given by
$\phi_0=0$ ($m_b=0$), while in the
confinement phase the minimum is shifted from zero
$\phi_0\not=0$ ($m_b\not=0$) which corresponds to
monopole condensation.  Thus the dual abelian gauge theory
with an action
$S[b]$ is equivalent to (the London limit of) the dual GL
theory (or the dual Abelian Higgs model with radial part of
the Higgs field being frozen), 
\begin{eqnarray}
  S[b] = \int d^4x \left[ 
 {-1 \over 4} b_{\mu\nu} b^{\mu\nu}
  +  | (\partial_\mu - ig^{-1} b_\mu ) \phi|^2
  + \lambda (|\phi|^2 - \phi_0^2)^2 + \cdots \right] ,
\end{eqnarray}
where we have rescaled the field 
$b_\mu \rightarrow b_\mu/g$.  Note that the inverse
coupling $g^{-1}$ has appeared as a coupling constant.  
This implies that the dual theory is suitable for
describing the strong coupling region.

\par
Now we compare our approach with the
previous approach
\cite{ST78,BS78} where the summation over the monopole
trajectories are performed.  
The monopole trajectories is expressed by the four-vector,
$
 x^\mu  = \bar x^\mu_l(\tau_l), \ l = 1, 2, \cdots , N
$
where $\tau_l$ is an arbitrary parameter characterizing the
trajectory  and $N$ is the total number of loops.
Then the monopole current is written as
\begin{eqnarray}
  K^\mu (x) = {4\pi \over g} \sum_{l=1}^N n_l \int d\tau_l
\
  \dot {\bar x}^\mu_l(\tau_l) 
  \delta^{(4)}(x-\bar x_l(\tau_l)) ,
\quad
 \dot {\bar x}^\mu := {\partial \bar x^\mu \over
\partial \tau} ,
\end{eqnarray}
with $n_l$ being the winding number.
Then the interaction $b_\mu K^\mu$ between
the dual field and the monopole current is written as
\begin{eqnarray}
  \int d^4x b_\mu(x) K^\mu (x) 
  = {4\pi \over g} \sum_{l=1}^N n_l
\int d\tau_l \ b_\mu(\bar x_l(\tau_l))
  \dot {\bar x}^\mu_l(\tau_l) .
\end{eqnarray}
Summation over all configurations containing arbitrary
number of monopole loops with all possible winding number
and trajectories is performed based on the identity
\cite{BS78},
\begin{eqnarray}
  &&\sum_{N=0}^\infty {1 \over N!} \int \prod_{l=1}^N 
  [d\bar x_l] \exp \left\{ i \sum_{l=1}^N \int d\tau_l 
  \left[ M \sqrt{\dot {\bar x}_l{}^2(\tau_l)} +
Q_\mu(\bar x_l(\tau_l)) 
  \dot {\bar x}^\mu_l(\tau_l) \right] \right\}
  \nonumber\\
  &&= \exp \Tr \ln H = \det (H)
  \nonumber\\
  &&=  \int [d\phi] \exp \left\{ i \int d^4 x \left[
  |(\partial_\mu + i Q_\mu(x)) \phi(x)|^2 - M^2 |\phi(x)|^2
\right] \right\} ,
  \nonumber\\
  &&  H := {1 \over 2}(p_\mu - Q_\mu)^2 - {1 \over 2} M^2 ,
\end{eqnarray}
where $\phi$ is a complex scalar field.  Both side are
equal to the vacuum-to-vacuum transition amplitude of the
theory consisting of charged scalar particles of mass $M$
in the presence of an external electromagnetic field
$Q_\mu$.
\par
If
$n_l$ is restricted to
$n_l=\pm 1$,  
the field theoretical quantity is obtained,
$$
 |(\partial_\mu + i g_m b_\mu(x)) \phi(x) |^2,
\quad g_m := {4\pi \over g}n ,
$$
where $Q_\mu = g_m b_\mu$  and
$\phi$  plays the role of the monopole.  
Assuming a mass term of the monopole field and the
repulsive self-interaction among the monopoles,
the low energy (infrared) effective theory of the
GL type, the effective dual GL theory, was proposed
\cite{Suzuki88},
\begin{eqnarray}
 {-1 \over 4}  b_{\mu\nu}(x) b^{\mu\nu}(x)
   + |(\partial_\mu + i g_m b_\mu(x)) \phi(x) |^2
 - \lambda (|\phi(x)|^2 - v^2)^2 .
\end{eqnarray}
If the monopole condensation occurs in the sense that
$|\phi(x)| \equiv v \not= 0$, the mass term of the dual
gauge field is generated, and the GL theory reduces to  
(note that the  normalization for the field
$b_\mu$ is different from (\ref{actionb}). )
$$
 {-1 \over 4}  b_{\mu\nu}(x) b^{\mu\nu}(x)
   + {1 \over 2} m_b^2 b_\mu(x)^2 , 
   \quad m_b \equiv g_m v ,
$$
This is the  so-called dual Meissner effect.
Precisely speaking, the classical solution 
\cite{NO73} $\phi(x)$ is not
a constant and is a function of $x$ such that 
$\phi(x) \rightarrow v$ as $|x| \rightarrow \infty$
and 
$\phi(x) \rightarrow 0$ as $|x| \rightarrow 0$.
The characteristic length separating both behaviors is the
coherence length $\xi:=\sqrt{2}/m_\phi$.  The ratio 
\begin{eqnarray}
\kappa_{GL} := \delta/\xi = m_\phi/(\sqrt{2}m_b)
\end{eqnarray}
is called the GL parameter where 
$\delta:=1/m_b$ is the penetration depth. 
The constant $|\phi(x)| \equiv v \not= 0$ corresponds to
$m_\phi=\infty$ or $\xi = 0$ and $\kappa_{GL}=\infty$,
a special case of type II superconductor
$\kappa_{GL}>1/\sqrt{2}$.
In the APEGT, this effect is expressed by the
$x$-dependent mass $m_b(x)$.

\subsection{Monopole action}
\par
It is easy to show that monopole
condensation actually occurs, if we use the lattice
version \cite{SS91,SS94} of the monopole action
\footnote{On the lattice, the monopole action is obtained
from the radially fixed Abelian Higgs model (of Villain
type) by lattice duality transformation \cite{BMK77}.
}
(\ref{maction}),
\begin{eqnarray}
  S_m  =  
    - {1 \over 4g^2}  \sum_{x} f_{\mu\nu}(x) f^{\mu\nu}(x)
 + \sum_{x,y} {1 \over g^2} 
 k^\mu(x) D_{\mu\nu}(x-y) k^\nu(y)   .
 \label{mlaction}
\end{eqnarray}
The monopole condensate (\ref{op}) is calculated as
follows.   From (\ref{mlaction}), we can
extract the self-mass term,
\begin{eqnarray}
 S_{ma} =   {D(0) \over g^2} \sum_{x} k^\mu(x) k^\mu(x) ,
 \quad D(0) < \infty .
\end{eqnarray}
The self-mass term with constant $|k_\mu(x)|=1$
(see \cite{SS91})
is proportional to length of monopole loops. The
probability that the monopole loop with length
$L$ will appear somewhere is
\begin{eqnarray}
 P_L = 7^L \exp (-S_{ma})
 = \exp \left[ (C - D(0)/g^2) L \right] ,
\end{eqnarray}
where $C= \ln 7$ for non backtracking walk on the
4-dimensional hypercubic lattice.
For sufficiently large $g^2$ $(g^2>D(0)/C)$, 
$P_L \uparrow \infty$ as $L \uparrow \infty$ and long
loops give dominant contribution to the functional
integral. 
On the other hand, 
$P_L \downarrow 0$ as $L \uparrow \infty$,
if $g^2$ is small $(g^2<D(0)/C)$. This indicates that
in the infinite volume limit long monopole loops make no
finite contribution.
This is a simple
energy-entropy (action-entropy) argument.
Taking into account the entropy contribution is equivalent
to adding an action,
\begin{eqnarray}
  S_{en}  = - C \sum_{x}    k^\mu(x) k^\mu(x), 
  \quad C < \infty .
\end{eqnarray}
Therefore we obtain
\begin{eqnarray}
  \Phi = \left( C - {D(0) \over g^2} \right)^{-1} .
\end{eqnarray}
This shows that, if the coupling $g$ is sufficiently
strong, we have positive $\Phi$ and non-zero $m_b$. 
In other words, if the entropy of a monopole
loop exceeds the energy, monopole condensation occurs. 
The region exhibiting monopole condensation extends to
smaller and smaller values of
$g$ for longer  loops due to recent studies
\cite{SS94}.
The above argument is valid for long loops.
For more details, see \cite{SS94}.  
The monopole action in the continuum needs more careful
treatment as in three-dimensional case \cite{Polyakov77}
which will be treated in a subsequent paper.
\par
In the usual language of field theory, the term
$k^\mu(x) D_{\mu\nu}(x-y) k^\nu(y)$
corresponds to the quartic self-interaction, especially,
the self-mass term 
$k^\mu(x)  k^\nu(x)$
to the contact quartic self-interaction. 
\footnote{
We remember that quartic self-interaction in the scalar
$\lambda \varphi^4$ theory can be understood as the
intersection probability of two random walks with
repulsive interaction. }
Therefore,
it is assumed that the self-interaction among monopole
loops does not essentially change the above picture.  It
should be remarked that higher order expansion generates
interactions between monopole loops.   For example, the
self-interaction among the monopoles, 
\begin{eqnarray}
\langle K_\mu(x)  K_\nu(y) K_\rho(z) K_\sigma(w)  \rangle
&=& \lambda(g) [\delta_{\mu\nu} \delta_{\rho\sigma}
\delta^{(4)}(x-y)\delta^{(4)}(z-w)\delta^{(4)}(x-z)
\\&&
+ \delta_{\mu\rho} \delta_{\nu\sigma} \delta^{(4)}(x-z)
\delta^{(4)}(y-w) \delta^{(4)}(x-y)
\\&&
+ \delta_{\mu\sigma} \delta_{\nu\rho} \delta^{(4)}(x-w)
\delta^{(4)}(y-z) \delta^{(4)}(x-y)] + \cdots ,
\end{eqnarray}
induces quartic self-interactions for $b_\mu$,
\begin{eqnarray}
&& \int d^4x d^4y d^4z d^4w
b_\mu(x) b_\nu(y) b_\rho(z) b_\sigma(w)
\langle K_\mu(x)  K_\nu(y) K_\rho(z) K_\sigma(w)  \rangle
\\
&=& 3 \lambda(g) \int d^4 x (b_\mu(x) b_\mu(x))^2 
+ \cdots .
\end{eqnarray}
This renormalizes the mass term in (\ref{actionb}) through
radiative corrections.  In this sense the criterion
(\ref{op}) is the tree-level criterion.
The monopole interaction is expected to be weak repulsive.

\subsection{Confinement and instanton}

The effective abelian theory $S[a]$ written in terms of
$a_\mu$ is obtained  by integrating out the dual gauge
field. This theory with an action $S[a]$
gives a dual description of the same physics as that given
by $S[b]$.
 Following the Zwanziger formalism \cite{Zwanziger71}, 
(we don't repeat the details, see \cite{Suzuki88} and
\cite{SST95}), if the dual gauge field acquires non-zero
mass $m_b$ (due to monopole condensation), we obtain  
\begin{eqnarray}
{\cal S}_{EFF}[a] &=& \int d^4x \left[
 {-1 \over 4g(\mu)^2}  f_{\mu\nu}(x) f^{\mu\nu}(x)
   + {1 \over 2} a^\mu(x) {n^2 m_b^2(x) \over 
   (n \cdot \partial)^2+n^2 m_b^2(x)} X_{\mu\nu}
(\partial)  a^\nu(x) \right] ,
   \nonumber\\
   && X_{\mu\nu}(\partial) := {1 \over n^2} 
   \epsilon^{\lambda\mu\alpha\beta}
   \epsilon^{\lambda\nu\gamma\delta} n_\alpha n_\gamma
   \partial_\beta \partial_\delta ,
   \label{actiona}
\end{eqnarray}
where $n$ is an arbitrary fixed four-vector appearing in
the Zwanziger formalism.  The coupling constant $g(\mu)$ is
the  running coupling constant obeying the same $\beta$
function as the YM theory.  In the limit
$m_b \rightarrow 0$, (\ref{actiona}) reduces to
(\ref{action1}).  The effective theory (\ref{actiona})
leads to the linear potential
$V(r) = \sigma r$
 between static color charges and the string tension
$\sigma$ is given by 
\begin{eqnarray}
   \sigma = {Q^2 \over 4\pi} m_b^2 f(\kappa_{GL}) ,
\end{eqnarray}
where $f(x)$ is a function depending on the method of
calculations \cite{Suzuki88,SST95}.
The essential part
$m_b^2$ in the string tension follows simply due to the
dimensional analysis, irrespective of the details of the
calculation. 
\par
The monopole condensation can be estimated based on the
classical configuration of ${\cal A}(x)$ satisfying the
gauge fixing condition $F^a[A,a]=0$,
\begin{eqnarray}
\langle K_\mu(x)  K_\mu(y)  \rangle
 &=& Z_{YM}^{-1} \int [d{\cal A}(x)] e^{-S_{YM}[{\cal A}]}
  \delta(F[A,a])
  \nonumber\\
&& \times \int [dK_\mu] \delta(K_\mu - {1 \over 2} g^2 
  \epsilon_{\mu\nu\rho\sigma} \partial^\nu (
 \epsilon^{ab3} A_\rho^a A_\sigma^b))  
  K_\mu(x)  K_\mu(y) .
  \label{mcexp}
\end{eqnarray}
Note that the MAG condition $F[A,a]=0$ is satisfied by the
classical multi-instanton solution \cite{Rajaraman,SS96} of
'tHooft type,
\begin{eqnarray}
  A_\mu^a(x) = 
 \bar \eta^a_{\mu\nu} \partial_\nu f(x),
 \quad
 \bar \eta^a_{\mu\nu} := \epsilon_{a\mu\nu} 
 + \delta_{a\mu}\delta_{\nu4} - \delta_{a\nu}\delta_{\mu4}
 = - \bar \eta^a_{\nu\mu} .
\end{eqnarray}
Therefore the
classical instanton configuration may have a possibility
to generate the monopole condensation.  Actually, it has
been shown that monopole loop formation and its
condensation are intimately correlated with the instanton
configuration
\cite{BOT96,CG95,HT96,BS96,STSM95,FSST97,FMT97,LRQ96}. 
Therefore it is quite interesting to clarify whether the
instanton configuration gives sufficient monopole loop
condensation for quark confinement. 
\par
The one-instanton solution has the form,
\begin{eqnarray}
  A_\mu^a(x)  = 2 \bar \eta^a_{\mu\nu} x_\nu g(x^2) ,
\end{eqnarray}
where the prime denotes the differentiation with respect
to the squared Euclidean distance 
$x^2=|x|^2=\sum_{\mu=1}^4 (x_\mu)^2$.
Then the monopole current is written as
\begin{eqnarray}
  K^\mu(x) &=&  \epsilon_{\mu\nu\alpha\beta}
\eta^3_{\alpha\beta} x^\nu  (x^2 g^2(x^2))',
\end{eqnarray}
where we have used the property,
\begin{eqnarray}
 \epsilon_{ABC} \bar \eta^B_{\mu\nu}
\bar \eta^C_{\alpha\beta}
 = \delta_{\mu\alpha} \bar \eta^A_{\nu\beta}
 - \delta_{\mu\beta} \bar \eta^A_{\nu\alpha}
 + \delta_{\nu\beta} \bar \eta^A_{\mu\alpha}
 - \delta_{\nu\alpha} \bar \eta^A_{\mu\beta} .
\end{eqnarray}
One-instanton solution with center at $x=z$ in the
singular gauge is given by 
\begin{eqnarray}
  g((x-z)^2) = {\rho^2 \over |x-z|^2(|x-z|^2+\rho^2)} ,
\end{eqnarray}
while in the non-singular gauge
\begin{eqnarray}
  g((x-z)^2) = {1 \over |x-z|^2+\rho^2} .
\end{eqnarray}
The expectation value (\ref{mcexp}) is replaced by the
integration over the collective coordinates, $\rho, z_\mu$.
Our preliminary calculation using the one-instanton 
solution in the singular gauge leads to non-zero monopole
loop condensation.  
According to \cite{BOT96}, however, one needs interpolating
gauge between the singular and non-singular to
derive monopole loop around the instanton.
Moreover, in order to
incorporate the interaction between
 instanton and anti-instanton and the resulting large
monopole loop formation 
\cite{BOT96,BS96,STSM95,FSST97,FMT97}, we
need more hard works.  The details of this problem
will be given in a subsequent paper
\cite{KMST97}.

\section{Inclusion of fermion}
\setcounter{equation}{0}
In order to discuss the QCD, we add the fermionic action,
\begin{eqnarray}
 S_F = \int d^4x \
\bar \psi [ i \gamma^\mu {\cal D}_\mu[{\cal A}]-m] \psi ,
\quad 
{\cal D}_\mu[{\cal A}] := \partial_\mu - i {\cal A}_\mu.
\end{eqnarray}
The contribution from the fermionic action is evaluated as
\begin{eqnarray}
&&  \int [d\bar \psi][d\psi] \exp \left\{ - \int d^4x
\bar \psi [ i \gamma^\mu {\cal D}_\mu[{\cal A}]-m] \psi
\right\}
\nonumber\\
&=& (\det [i \gamma^\mu {\cal D}_\mu[{\cal A}]-m])^{N_f}
\nonumber\\
&=& \exp \left[
 N_f \ln \det [i \gamma^\mu {\cal D}_\mu[{\cal A}]-m] \right]
\nonumber\\
&=& \exp \left[ {N_f \over 2} 
\ln \det [i \gamma^\mu {\cal D}_\mu[{\cal A}]-m]^2 \right] .
\end{eqnarray}
In a similar way as in section 2, we can calculate the 
logarithmic determinant,
\begin{eqnarray}
&& \Tr(\exp [-t (i \gamma^\mu {\cal D}_\mu[{\cal A}])^2] )
-  \Tr(\exp [-t (i \gamma^\mu \partial_\mu)^2] )
\nonumber\\
&=&  \int d^4x {g^2 \over 16\pi^2} {2 \over 3}r(F)  
({\cal F}_{\mu\nu}^a)^2  + O(t) ,
\end{eqnarray}
and
\begin{eqnarray}
\ln {\det (i \gamma^\mu D_\mu[{\cal A}])^2 \over  
\det (i \gamma^\mu \partial_\mu)^2}
= \int d^4x  {g^2 \over 16\pi^2} {2 \over 3}r(F) \ln \mu^2 
({\cal F}_{\mu\nu}^a)^2 ,
\end{eqnarray}
where $r(F)$ is the dimension of fermion representation. In
this calculation, we have used the commutator,
\begin{eqnarray}
 [{\cal D}_\mu[{\cal A}], {\cal D}_\nu[{\cal A}]] 
= - i {\cal F}_{\mu\nu} .
\end{eqnarray}
\par
At one-loop level, it is easy to see that we can replace
$({\cal F}_{\mu\nu}^a)^2$ in this contribution by
$(f_{\mu\nu})^2$.
If we add this contribution to the
APEGT obtained in section 2, the APEGT of QCD is obtained
(apart from the gauge-fixing term and the abelian ghost
term),
\begin{eqnarray}
  S &=&  \int d^4x \Big[
    - {1+z_a \over 4g^2}  f_{\mu\nu} f^{\mu\nu}
    - {1+z_b \over 4} g^2 b_{\mu\nu} b^{\mu\nu}
 - z_c b_\mu K^\mu 
\nonumber\\
&&+ \bar \psi ( i \gamma^\mu {\cal D}_\mu[a]-m) \psi 
+ i \bar c^a D_\mu^{bc}[a] D_\mu^{bc}[a] c^c 
\Big] ,
\nonumber\\
{\cal D}_\mu[a] &:=& \partial_\mu - i a_\mu T^3 .
\label{APEGTQCD}
\end{eqnarray}
In the region $K_\mu \cong 0$, it is clear that this theory
recovers the one-loop beta function of QCD,
\begin{eqnarray}
 b_0 = {11 \over 3} C_2(G) - {4 \over 3} N_f r(F) .
\end{eqnarray}
The monopole condensation and resulting dual Meissner
effect can be treated in a similar way to section 4.
We can discuss the chiral symmetry breaking based on
APEGT of QCD (\ref{APEGTQCD}), see e.g. \cite{SST95}.

\section{Lower dimensional case}
\setcounter{equation}{0}

In the 2+1 dimensional case, we introduce the auxiliary
vector field $B_\mu$ (instead of the tensor field
$B_{\mu\nu}$ in 3+1 dimensional case).  Then,
corresponding to (\ref{apBFYM}) or (\ref{apYM}), the action
is rewritten as
\begin{eqnarray}
   S_{apBFYM}[{\cal A}, B] =  \int d^3x \left[
     {1 \over 4} \epsilon^{\mu\nu\rho} B_{\rho} 
     (f_{\mu\nu} + {\cal C}_{\mu\nu}) 
     - {1 \over 4} g^2 B_{\mu} B^{\mu} 
    - {1 \over 4g^2} ({\cal S}_{\mu\nu}^a)^2 \right] ,
    \label{apBFYM3d}
\end{eqnarray}
or
\begin{eqnarray}
   S_{apYM}[{\cal A}, B] &=&  \int d^3x \Big[
     - {1 \over 4g^2} 
     (f_{\mu\nu}f_{\mu\nu} + 2 f_{\mu\nu}{\cal C}_{\mu\nu}) 
  +  {1 \over 4} \epsilon^{\mu\nu\rho} B_{\rho} 
   {\cal C}_{\mu\nu}
     - {1 \over 4} g^2 B_{\mu} B^{\mu} 
\nonumber\\&&
    - {1 \over 4g^2} ({\cal S}_{\mu\nu}^a)^2 \Big] .
    \label{apYM3d}
\end{eqnarray}
At the tree level, the dual vector field has the
respective correspondence,
\begin{eqnarray}
  B_{\mu}  \leftrightarrow   {1 \over 2}
\epsilon^{\mu\rho\sigma} (f_{\rho\sigma}
+{\cal C}_{\rho\sigma}),
\quad
 {1 \over 2}
\epsilon^{\mu\rho\sigma} {\cal C}_{\rho\sigma} .
\end{eqnarray}
In order to discuss the monopole contribution, we use the
decomposition,
\footnote{
The vector $B_\mu$ has  has three degrees of freedom, while
the real scalar $\phi$ has one and the vector $\chi_\mu$ has
three. One redundant degrees of freedom corresponds to that
of the gauge transformation of $\chi_\mu$.
}
\begin{eqnarray}
  B_{\mu} = \partial_\mu \phi + {1 \over 2}
\epsilon_{\mu\alpha\beta} \chi^{\alpha\beta}  ,
\quad
 \chi_{\mu\nu} &:=& \partial_\mu \chi_\nu - \partial_\nu
\chi_\mu .
 \label{Hd3d}
\end{eqnarray}
Hence APEGT of the 2+1 dimensional YM theory is given by
\begin{eqnarray}
   S_1[a,\phi,\chi] &=&  \int d^3x \Big[
     - {1 \over 4g^2} 
     f_{\mu\nu}f_{\mu\nu}  
     - {1 \over 4} g^2 \left[
 (\partial_\mu \phi)^2    + \chi_{\mu\nu}^2 \right]
\Big] , 
    \label{apYM3d1}
\end{eqnarray}
and
\begin{eqnarray}
     Q_{\mu\nu}^{ab} &:=&  
   (D_\rho[a] D_\rho[a])^{ab} \delta_{\mu\nu}
   - 2 \epsilon^{ab3} f_{\mu\nu}
 + {1 \over 2} g^2 \epsilon^{ab3}
   (\epsilon_{\mu\nu\rho} \partial^\rho \phi
   + \chi_{\mu\nu})
\nonumber\\&&
 - 2i g^2 \xi (\bar c^a  c^b - \bar c^c c^c \delta^{ab})
\delta_{\mu\nu}
 - D_\mu[a]^{ac} D_\nu[a]^{cb} 
 + {1 \over \alpha}  D_\mu[a]_\xi^{ac} D_\nu[a]_\xi^{cb} .
\label{defKd3}
\end{eqnarray}
In 2+1 dimensional case, instead of interaction $b_\mu
K^\mu$ between the dual gauge field and the magnetic
current, we obtain the interaction term between the dual
scalar $\phi$ and the monopole density $\rho$, 
\begin{eqnarray}
 \rho(x) \phi(x) , \quad
 \rho(x) := \epsilon^{\mu\nu\rho} 
 \partial_{\rho} {\cal C}_{\mu\nu}(x) ,
\end{eqnarray}
since
\begin{eqnarray}
 \int d^3x \epsilon^{\mu\nu\rho} B_{\rho} {\cal C}_{\mu\nu}
&=&   \int d^3x \left[ - \phi \epsilon^{\mu\nu\rho}
\partial_{\rho} 
   {\cal C}_{\mu\nu}
   + \chi_{\mu\nu}{\cal C}_{\mu\nu} \right] .
\end{eqnarray}
The effective dual theory is the scalar theory with 
\begin{eqnarray}
   S[\phi]  &=&  
   \int d^3x (\partial_\mu \phi(x))^2 
+  \int d^3x \langle \rho(x) \rangle  \phi(x)
\nonumber\\&&
+ {1 \over 2} \int d^3x \int d^3y
\langle \rho(x) \rho(y) \rangle_c \phi(x) \phi(y) 
+ \cdots .
\end{eqnarray}

\par
In 1+1 dimensional case, the dual tensor reduces to a
one-component scalar $B$.
\begin{eqnarray}
   S_{apBFYM}[{\cal A}, \phi] =  \int d^2x \left[
     {1 \over 4} \epsilon^{\mu\nu}  
     (f_{\mu\nu} + {\cal C}_{\mu\nu}) \phi
     - {1 \over 4} g^2 \phi^2
    - {1 \over 4g^2} ({\cal S}_{\mu\nu}^a)^2 \right] ,
    \label{apBFYM2d}
\end{eqnarray}
or
\begin{eqnarray}
   S_{apYM}[{\cal A}, \phi] &=&  \int d^2x \Big[
     - {1 \over 4g^2} 
     (f_{\mu\nu}f_{\mu\nu} + 2 f_{\mu\nu}{\cal C}_{\mu\nu}) 
  +  {1 \over 4} \epsilon^{\mu\nu} \phi {\cal C}_{\mu\nu}
     - {1 \over 4} g^2 \phi^2
\nonumber\\&&
    - {1 \over 4g^2} ({\cal S}_{\mu\nu}^a)^2 \Big] .
    \label{apYM3dd}
\end{eqnarray}
The tree-level correspondence is given by
\begin{eqnarray}
  \phi  \leftrightarrow   {1 \over 2}
\epsilon^{\rho\sigma} (f_{\rho\sigma}
+{\cal C}_{\rho\sigma}),
\quad
 {1 \over 2}
\epsilon^{\rho\sigma} {\cal C}_{\rho\sigma} .
\end{eqnarray}
Thus 1+1 dimensional YM theory is reduced to an effective
abelian gauge theory with
\begin{eqnarray}
   S_1[a,\phi] &=&  \int d^2x \left[
     - {1 \over 4g^2} 
     f_{\mu\nu}f_{\mu\nu}  
      - {1 \over 4} g^2 \phi^2 \right], 
    \label{apYM2d1}
\end{eqnarray}
and
\begin{eqnarray}
     Q_{\mu\nu}^{ab} &:=&  
   (D_\rho[a] D_\rho[a])^{ab} \delta_{\mu\nu}
   - 2 \epsilon^{ab3} f_{\mu\nu}
 + {1 \over 2} g^2 \epsilon^{ab3}
   \epsilon_{\mu\nu} \phi 
\nonumber\\&&
 - 2i g^2 \xi (\bar c^a  c^b - \bar c^c c^c \delta^{ab})
\delta_{\mu\nu}
 - D_\mu[a]^{ac} D_\nu[a]^{cb} 
 + {1 \over \alpha}  D_\mu[a]_\xi^{ac} D_\nu[a]_\xi^{cb} .
\label{defKd2}
\end{eqnarray}
In this case, the interaction term is induced,
\begin{eqnarray}
    \phi(x) \epsilon_{\mu\nu} f_{\mu\nu}(x).
\end{eqnarray}
It is interesting to compare these formulations with the
previous approaches \cite{Polyakov77,CD77,RU78,Levine80}.
Detailed analyses of the lower
dimensional case will be given in a forthcoming paper.

\section{Conclusion and discussion}
\setcounter{equation}{0}

We have derived abelian-projected effective  gauge theories
(APEGT) of YM theory and QCD. This has been performed by
integrating out all off-diagonal non-Abelian gauge fields
belonging to $SU(2)/U(1)$.
The obtained APEGT is written in terms of
the maximal abelian gauge field
$a_\mu$ and the dual abelian gauge field
$b_\mu$ which couples to the magnetic monopole current
$K_\mu$.
First, we have shown that
the APEGT has the same one-loop beta function as the
original non-Abelian gauge theories.   Hence the APEGT
exhibits asymptotic freedom (at one-loop level).
\par
Next,  we have shown that the dual vector field introduced
to linearize the gluon self-interaction has an interaction
with the magnetic current.   Due to this interaction, the
dual gauge field can become massive if the monopole loop
condensation occurs. This is interpreted as the dual
Meissner effect.  We have shown that the mass of the dual
gauge field is given by the monopole loop condensation
$
 \langle K_\mu(x) K^\mu(x) \rangle/\delta^{(4)}(0)
 \not =0 .
$
This is our criterion of dual superconductivity.
A method of showing monopole condensation is to
consider the monopole action.  The lattice monopole action
\cite{SS91,SS94} gives a simple proof of monopole
condensation.
\par
If we apply the Zwanziger formalism to the APEGT with
magnetic monopole, we can show that the static quark
potential contains a linear part proportional to the
quark separation.   APEGT with monopole is sufficient
to show quark confinement. This supports the abelian
dominance. The monopole dominance will be confirmed by
evaluating the monopole condensate, since the string
tension is determined from the mass $m_b$ of dual gauge
field.  
We have pointed out that this condensation can
be estimated by the classical instanton configuration. 
Intimate relationship between confinement and instanton
will be understood from the viewpoint of topological field
theory of Schwarz type, BF-YM theory.
\par
This work justifies some aspects of
the pioneering works by Ezawa and Iwazaki \cite{EI82}
and Suzuki
\cite{Suzuki88} based on the effective dual GL model. 
However, the APEGT has no free parameter and is of
predictive power in sharp contrast with the previous works
where the abelian dominance was assumed from the beginning.
The APEGT has the complete correspondence to the original
YM theory. 
\par
We have chosen the gauge group SU(2) for mathematical
simplicity. To discuss the confinement in the real world, we
must discuss the SU(3) case.  This case will be treated
in a subsequent paper
\cite{KMST97}.

\section*{Acknowledgments}
This work was inspired by a series of lectures given
by Tsuneo Suzuki  at Chiba University in January
1997.  After submitting this paper for publication, I have
enjoyed fruitful discussions with  Tadahiko Kimura,
Taro Kashiwa, Shoichi Sasaki, Hideo Suganuma and Hiroshi
Toki. I would like to thank all of them.  Especially, many
suggestions by  H. Suganuma greatly helped
me to improve the paper. 
I also thanks  Maxim Chernodub for sending valuable
comments.  This work is supported in part by the
Grant-in-Aid for Scientific Research from the Ministry of
Education, Science and Culture (No.07640377).

\appendix
\section{APEGT of BF-YM theory}
\setcounter{equation}{0}

In the similar way as in section 2, the APEGT of BF-YM is
obtained as
\begin{eqnarray}
  S_0 + S_1 + S_2
&=& 
\int d^4x \Big[
- z_a {1 \over 4g^2} f_{\mu\nu}f^{\mu\nu}
    - {1 \over 4} (1+z_b) g^2 B_{\mu\nu} B^{\mu\nu}
+  {1 \over 2} (1-z_c)  B_{\mu\nu}  \tilde f_{\mu\nu}  
\nonumber\\&&
+ i \bar c^a D^\mu{}^{ac}[a]^\xi D_\mu^{cb}[a] c^b
\Big] ,
\end{eqnarray}
where 
\begin{eqnarray}
 z_a = - {2 \over 3} \kappa {g^2 \over 16\pi^2} \ln \mu, 
\quad
 z_b = + 2 \kappa {g^2 \over 16\pi^2} \ln \mu,
\quad
 z_c = + 2 \kappa {g^2 \over 16\pi^2} \ln \mu .
\end{eqnarray}
Integrating out the tensor field $B$, we obtain
\begin{eqnarray}
  S_E
&=& 
\int d^4x \Big[
- z_a {1 \over 4g^2} f_{\mu\nu}f^{\mu\nu}
- {1 \over 4g^2}(1+z_b)^{-1} (1-z_c)^2 
f_{\mu\nu}f^{\mu\nu}
\nonumber\\&&
+  i \bar c^a D^\mu{}^{ac}[a]^\xi D_\mu^{cb}[a] c^b
\Big] .
\end{eqnarray}
Hence, at one-loop level, this reduces to
\begin{eqnarray}
  S_E
&=& 
\int d^4x \left[
- (1+h) {1 \over 4g^2} f_{\mu\nu}f^{\mu\nu}
+ i \bar c^a D^\mu{}^{ac}[a]^\xi D_\mu^{cb}[a] c^b
\right] ,
\end{eqnarray}
where
\begin{eqnarray}
 h = z_a - z_b - 2z_c 
= - {20 \over 3} \kappa {g^2 \over 16\pi^2} \ln \mu .
\end{eqnarray}
This agrees with the APEGT of YM theory given in section 2.
Therefore, two types of APEGT are equivalent to each other.

\section{Ghost interaction and gauge fixing}
\setcounter{equation}{0}

If we adopt more general gauge fixing functional,
\begin{eqnarray}
  G &=& \sum_{\pm}
  \bar c^{\mp} (F^{\pm}[A,a] + {\alpha \over 2} \phi^{\pm})
  + \bar c^3 (F^{3}[a] + {\beta \over 2} \phi^{3}) 
  \nonumber\\&&
  + \alpha \eta \sum_{\pm} (\pm) 
  \bar c^3 \bar c^{\pm} c^{\mp}
  + \alpha \zeta c^3 \bar c^+ \bar c^- .
\label{G2}
\end{eqnarray}
the gauge fixing part 
${\cal L}_{GF}=-i \delta_B G$ has the additional
contribution,
\begin{eqnarray}
 {\cal L}'_{GF}
 &=& - \sum_{\pm} (\pm) \bar c^{\mp} {\alpha \over \beta}
\eta F^3[a] c^{\pm}
- \sum_{\pm} (\pm) \bar c^3 \eta F^{\pm}[A,a] c^{\mp}
+ \sum_{\pm} (\pm) \bar c^{\mp} \zeta F^{\pm}[A,a] c^3
\nonumber\\&&
- \alpha (1+\zeta) \eta \sum_{\pm} \bar c^3 c^3 \bar
c^{\pm} c^{\mp}
- \alpha (\zeta + {\alpha \over \beta} \eta^2) 
\bar c^+ \bar c^- c^+ c^- .
\end{eqnarray}
Therefore, the U(1) invariant four-ghosts interaction term
$\bar c^+ \bar c^- c^+ c^-$ coming from the
expansion of $\ln \det Q$, 
\begin{eqnarray}
 (\bar c^a  c^b - \bar c^c c^c \delta^{ab})
(\bar c^b  c^a - \bar c^d c^d \delta^{ba})
= - 2 \bar c^1 c^1 \bar c^2 c^2
= - 2 \bar c^+ \bar c^- c^+ c^- 
\end{eqnarray}
is canceled by adding
the BRST exact term,
$-i\delta_B(c^3 \bar c^+ \bar c^-) 
= - i \{ Q_B, c^3 \bar c^+ \bar c^- \}$.
Such a term does not influence the physical state
characterized by 
$Q_B |phys \rangle = 0$.
This is an implication of the renormalizability of YM
theory in MAG.

\section{Magnetic monopole and Dirac string in SU(2) gauge
theory}
\setcounter{equation}{0}

In this appendix, we discuss how the abelian objects, Dirac
magnetic monopole and Dirac string, are produced due to
singular gauge transformation in SU(2) non-Abelian gauge
theory.
\footnote{This appendix is deeply indebted 
to H. Suganuma and H. Ichie \cite{SI97}.
}

\par
The Non-Abelian field strength ${\cal F}_{\mu\nu}$ is
defined using the covariant derivative,
\begin{eqnarray}
  {\cal D}_\mu := \partial_\mu - ig {\cal A}_\mu 
\end{eqnarray}
as
\begin{eqnarray}
 {\cal F}_{\mu\nu} = {i \over g}
 [ {\cal D}_\mu, {\cal D}_\nu ]
 = {i \over g} 
 [ \partial_\mu - ig {\cal A}_\mu , 
 \partial_\nu - ig {\cal A}_\nu  ] .
\end{eqnarray}
This is rearranged as
\begin{eqnarray}
 {\cal F}_{\mu\nu} &=& {i \over g} 
 [ \partial_\mu , \partial_\nu  ]
 + [ \partial_\mu , {\cal A}_\nu  ]
 - [ \partial_\nu , {\cal A}_\mu ]
  - ig [{\cal A}_\mu ,  {\cal A}_\nu  ]
  \nonumber\\
  &=& {i \over g} 
 [ \partial_\mu , \partial_\nu  ]
 +  \partial_\mu  {\cal A}_\nu  
 -   \partial_\nu   {\cal A}_\mu 
  - ig [{\cal A}_\mu ,  {\cal A}_\nu  ] .
  \label{fss}
\end{eqnarray}
It should be remarked that the first term on the RHS in the
final line  can not be neglected when there is a singularity.
We consider the local gauge transformation, 
\begin{eqnarray}
 {\cal A}_{\mu}  \rightarrow 
 {\cal A}_{\mu}' := U {\cal A}_{\mu} U^\dagger 
 + {i \over g} U \partial_\mu U^\dagger .
 \label{gt}
\end{eqnarray}
Straightforward calculation using (\ref{gt}) leads to
\begin{eqnarray}
 {\cal F}_{\mu\nu}' &:=& 
 \partial_\mu  {\cal A}_\nu'  
 -   \partial_\nu   {\cal A}_\mu' 
  - ig [{\cal A}_\mu' ,  {\cal A}_\nu'  ]  
 \\
 &=& 
   U (\partial_\mu  {\cal A}_\nu  
 -   \partial_\nu   {\cal A}_\mu 
  - ig [{\cal A}_\mu ,  {\cal A}_\nu  ]) U^\dagger 
 + {i \over g} U
 [ \partial_\mu , \partial_\nu  ] U^\dagger .
\end{eqnarray}
This is consistent with (\ref{fss}), that is, the
field strength transforms covariantly,
\begin{eqnarray}
 {\cal F}_{\mu\nu}  \rightarrow 
 {\cal F}_{\mu\nu}' := U {\cal F}_{\mu\nu} U^\dagger .
\end{eqnarray}
In what follows we assume that ${\cal A}_{\mu}$ is not
singular and that the singularity in ${\cal A}_{\mu}'$
comes from the gauge rotation $U$.  In such a case, we
call $U$ the singular gauge rotation.
Therefore, the gauge-transformed field strength is composed
of two parts, the regular and the singular part,
\begin{eqnarray}
 {\cal F}_{\mu\nu}' &=& 
 {\cal F}_{\mu\nu}^{r}{}' + {\cal F}_{\mu\nu}^{s}{}',
 \nonumber\\
 {\cal F}_{\mu\nu}^{r}{}' 
 &:=& U (\partial_\mu  {\cal A}_\nu  
 -   \partial_\nu   {\cal A}_\mu 
  - ig [{\cal A}_\mu ,  {\cal A}_\nu  ]) U^\dagger  ,
  \nonumber\\
  {\cal F}_{\mu\nu}^{s}{}' 
  &:=& {i \over g} U
 [ \partial_\mu , \partial_\nu  ] U^\dagger .
\end{eqnarray}
\par
First, we show that only the second part of the potential
${\cal A}{'}_{\mu}(x)$,
\begin{eqnarray}
 {\cal A}^s_{\mu}(x)  :=  
   {i \over g} U(x) \partial_\mu U^\dagger(x)  
 \label{sp}
\end{eqnarray}
gives rise to the non-vanishing magnetic current.
The diagonal part $a_\mu^s$ of the gauge potential ${\cal
A}^s_{\mu}$
 is singular on the point where the Dirac string exists.
The direction of the Dirac string can be changed
arbitrarily by the gauge transformation. Hence the
Dirac string is not a physical object.  Actually, the
magnetic charge is shown to obey the Dirac quantization
condition.  This can be seen as follows. 
\par
The local SU(2) matrix $U(x)$ can be written in terms of
three Euler's angles $\alpha, \beta, \gamma$,
\begin{eqnarray}
 U(x) 
&=& e^{i \gamma(x) \sigma_3/2} 
   e^{i \beta(x) \sigma_2/2} 
   e^{i \alpha(x) \sigma_3/2}  
   \nonumber\\
 &=& \pmatrix{
 e^{{i \over 2}(\alpha(x)+\gamma(x))} 
 \cos {\beta(x) \over 2} &
  e^{{i \over 2}(\alpha(x)-\gamma(x))}
  \sin {\beta(x) \over 2} \cr
  - e^{-{i \over 2}(\alpha(x)-\gamma(x))}
  \sin {\beta(x) \over 2} & 
  e^{-{i \over 2}(\alpha(x)+\gamma(x))}
  \cos {\beta(x) \over 2} } .
   \label{U0}
\end{eqnarray} 
Using the residual U(1) invariance after MAG, we can
choose 
$\gamma(x) = - \alpha(x)$.
A convenient choice is to take
$\alpha(x)  = -\gamma(x) = \varphi(x)$, 
$\beta(x) = \theta(x)$
and identity the angle $\theta$ and $\varphi$ with the polar
and the azimuthal angles in the three-dimensional polar
coordinate of SU(2)
so that
\begin{eqnarray}
 U(x)_{\theta, \varphi}
 &=& \exp ({i \theta \vec e_\varphi \cdot \vec \sigma/2})
 = \pmatrix{ \cos {\theta(x) \over 2} &
  e^{i\varphi(x)} \sin {\theta(x) \over 2} \cr
  - e^{-i\varphi(x)} \sin {\theta(x) \over 2} & 
   \cos {\theta(x) \over 2} } 
 \nonumber\\
 &=& \cos {\theta(x) \over 2} + i \vec \sigma \cdot \vec
e_\varphi \sin {\theta(x) \over 2} ,
  \\ \vec e_\varphi &:=& - \sin \varphi \ \vec e_X 
 + \cos \varphi \ \vec e_Y ,
   \label{U}
\end{eqnarray} 
where $(X,Y,Z)$ is identified with the space coordinates of
$x^\mu = (0,\vec r) = (0,X,Y,Z)$ and
\begin{eqnarray}
  0 < \theta := \arctan {\sqrt{X^2+Y^2} \over Z} < \pi,
  \quad
  0 < \varphi := \arctan  {Y \over X} < 2\pi .
\end{eqnarray} 
This choice does not lose generality, since we can always
rotate the matrix using the residual U(1) degrees of
freedom, see \cite{CG95} for details.
\footnote{
If we take $\gamma(x)=\alpha(x)$ and write
$\alpha(x)  = \gamma(x) = \varphi(x)$, 
$\beta(x) = \theta(x)$,
\begin{eqnarray}
 U(x) 
= \pmatrix{
 e^{i \varphi(x)} 
 \cos {\theta(x) \over 2} &
    \sin {\theta(x) \over 2} \cr
  -   \sin {\theta(x) \over 2} & 
  e^{- i \varphi(x)}
  \cos {\theta(x) \over 2} } .
   \label{U2}
\end{eqnarray} 
For this choice of $\gamma$, the Dirac string appears on
the positive $Z$ axis, since the $\beta=0, \pi$ corresponds
to 
\begin{eqnarray}
 U(x)_{0,\varphi} 
 = \pmatrix{
 e^{i \varphi(x)}   &     0 \cr
     0 &  e^{- i \varphi(x)} } ,
\quad
 U(x)_{\pi,\varphi} = \pmatrix{
 0   &    1  \cr  - 1  &  0 } .
\end{eqnarray} 
}
\par
For the gauge rotation (\ref{U}), the three-dimensional
part of ${\cal A}^s_\mu$ is
\begin{eqnarray}
 \vec {\cal A}^s(x)  
&=& {1 \over gr}(\cos \varphi(x) \ \vec e_\varphi + \sin
\varphi(x)
\ \vec e_\theta) T^1 
+ {1 \over gr}(\sin \varphi(x) \ \vec e_\varphi - \cos
\varphi(x)
\ \vec e_\theta) T^2
\nonumber\\&&
+ {1 \over g r} \tan{\theta(x) \over 2}
  \vec e_\varphi   T^3,
\end{eqnarray}
where we have used
\begin{eqnarray}
  \nabla :=  \vec e_r {\partial \over \partial r}
  + {\vec e_\theta \over r}{\partial \over \partial \theta}
  + {\vec e_\varphi \over r \sin \theta} 
  {\partial \over \partial \varphi} .
\end{eqnarray}
\par
The  diagonal  abelian part is defined by    
\begin{eqnarray}
 a_{\mu}' := 2 \tr(T^3 {\cal A}_{\mu}') .
\end{eqnarray}
In this case,
\footnote{
The 4-dimensional expression is given by
\begin{eqnarray}
  a^s_\mu(x) =
  - {1 \over g} [\cos \beta(x) \partial_\mu \alpha(x) +
\partial_\mu \gamma(x) ] .  
\end{eqnarray}
The angle $\gamma(x)$ does not appear in the U(1) invariant
quantity. Actually, the magnetic current given by
\begin{eqnarray}
  k_\mu(x) = {1 \over g} \epsilon_{\mu\nu\rho\sigma}
\partial_\nu [\partial_\rho \cos \beta(x) \partial_\sigma
\alpha(x)]    
\end{eqnarray}
does not contain the angle $\gamma$.
For more details, see \cite{BOT96}.
}
\begin{eqnarray}
  \vec a^s(x) 
  = {1 \over g r} \tan{\theta \over 2}
  \vec e_\varphi  
   = {1 \over g r}{1-\cos \theta \over \sin \theta}
  \vec e_\varphi ,  
\end{eqnarray}
or
\begin{eqnarray}
   a_s^\mu(x) = (a_0^s(x), \vec a^s(x))
   = {1 \over g r(r+Z)} (0, -Y, X, 0) .
\end{eqnarray}
The vector potential $\vec a^s$ is singular on the negative
$Z$ axis and is not defined for $\theta=\pi$. Then the
rotation is given by
\begin{eqnarray}
   \nabla \times \vec a^s(x) = \vec B_m + \vec B_{DS}
   =  {\vec r \over gr^3}
   +  {4\pi \over g} \delta(X) \delta(Y) \theta(-Z) \vec
e_Z .
\end{eqnarray}
This implies that 
$\nabla \times \vec a^s(x)  = {\vec r \over gr^3}$
except along the negative $Z$ axis.
The singularity along the negative $Z$ axis is called the
Dirac string.  This can not be avoided as long as one uses
the single expression for the gauge potential in the whole
space.  A method to  avoid the singularity is using the
Wu-Yang monopole
\cite{WY75}.  It is impossible to construct a single
singularity-free potential which is defined everywhere.
When considering the total space, we need at least two
expressions for the vector potential.

The magnetic monopole
sits at $\vec r=0$,
\begin{eqnarray}
   \nabla \cdot \vec B_m = k_0^m(x), \quad
   k_0^m(x) = {4\pi \over g} \delta^{(3)}(x) .
\end{eqnarray}
The four-dimensional expression of the magnetic current is
\begin{eqnarray}
   {1 \over 2} \epsilon_{\mu\nu\rho\sigma} 
   \partial_\sigma f_{\mu\nu} = k_\rho(x)  , \quad
 k_\mu(x) = {4\pi \over g} \delta_{\mu 0} \delta^{(3)}(x) .
\end{eqnarray}
where the abelian-projected field strength is defined,
\begin{eqnarray}
 f_{\mu\nu} 
:= \partial_\mu  a_\nu^s  -   \partial_\nu   a_\mu^s
 :=  \tr[T^3(\partial_\mu {\cal A}^s_{\nu}
 - \partial_\nu {\cal A}^s_{\mu})]
\end{eqnarray}
The magnetic flux $\Phi$ obtained by integrating $\vec B_m$
over any closed surface containing the origin is
\begin{eqnarray}
  \Phi_m := \int_S \vec B_m \cdot d\vec S 
  = {4\pi \over g} .
  \label{mg}
\end{eqnarray}
On the other hand, the magnetic flux $\Phi$ obtained by
integrating
$\vec B_{Ds}$ over any closed surface containing the
origin is
\begin{eqnarray}
  \Phi_{Ds} := \int_S \vec B_{Ds} \cdot d\vec S 
  = - {4\pi \over g} .
  \label{Ds}
\end{eqnarray}

\par
We observe that the singular gauge potential 
${\cal A}^s_{\mu}$ satisfies the following
relation,
\begin{eqnarray}
&& \partial_\mu {\cal A}^s_{\nu}
 - \partial_\nu {\cal A}^s_{\mu}
 \nonumber\\
 &=& {i \over g} \left\{ 
 \partial_\mu (U \partial_\nu U^\dagger) -
\partial_\nu (U \partial_\mu U^\dagger) \right\}
 \nonumber\\
 &=& {i \over g} \left\{ 
  (\partial_\mu U) (\partial_\nu U^\dagger) -
 (\partial_\nu U) (\partial_\mu U^\dagger) \right\}
 + {i \over g} \left( 
    U \partial_\mu \partial_\nu U^\dagger  -
  U \partial_\nu \partial_\mu U^\dagger  \right)
 \nonumber\\
 &=& {i \over g} \left\{ 
  (\partial_\mu U) U^\dagger U (\partial_\nu U^\dagger) -
 (\partial_\nu U)  U^\dagger U (\partial_\mu U^\dagger)
\right\}
 + {i \over g}   
  ( U [\partial_\mu , \partial_\nu] U^\dagger) 
 \nonumber\\
&=& {i \over g} \left\{ 
  -(U \partial_\mu U^\dagger) (U \partial_\nu U^\dagger) +
  (U \partial_\nu U^\dagger) (U \partial_\mu U^\dagger)
\right\}
 + {i \over g}   
  ( U [\partial_\mu , \partial_\nu] U^\dagger) 
 \nonumber\\
&=&  {i \over g}  
   [i U \partial_\mu U^\dagger, i U \partial_\nu U^\dagger]
 + {i \over g}   
  ( U [\partial_\mu , \partial_\nu] U^\dagger) 
 \nonumber\\
&=&   i g   
   [ {\cal A}^s_{\mu}, {\cal A}^s_{\nu}]
 + {i \over g}   
  ( U [\partial_\mu , \partial_\nu] U^\dagger) ,
  \label{id}
\end{eqnarray}
where we have used 
\begin{eqnarray}
  UU^\dagger = 1 ,
  \quad  \partial_\mu (UU^\dagger)
  = (\partial_\mu U) U^\dagger + U (\partial_\mu U^\dagger)
  = 0 .
\end{eqnarray}
Hence the abelian-projected field
strength reads
\begin{eqnarray}
 f_{\mu\nu} 
=    \tr(T^3 i g   
   [ {\cal A}^s_{\mu},  {\cal A}^s_{\nu}])
 + \tr(T^3 {i \over g}   
   U [\partial_\mu , \partial_\nu] U^\dagger) .
   \label{id2}
\end{eqnarray}
If $U$ was not singular, the last term in (\ref{id}) or
(\ref{id2}) was absent, since ${\cal A}^s_{\mu}$ is a pure
gauge which gives vanishing field strength for
non-singular $U(x)$,
\begin{eqnarray}
 {\cal F}_{\mu\nu} :=  \partial_\mu {\cal A}^s_{\nu}
 - \partial_\nu {\cal A}^s_{\mu}
  - i g [ {\cal A}^s_{\mu}, {\cal A}^s_{\nu}] \equiv 0 .
\end{eqnarray}
The last term in (\ref{id}) corresponds to the singularity
due to Dirac string as shown shortly. 
\par
Now we  clarify the physical meaning of the
last term,
$
 {i \over g} 
 (U[ \partial_\mu , \partial_\nu  ] U^\dagger)^{(3)} .
$
 We show that
 \footnote{
  This is derived also from Homotopy theory, 
  $
  \Pi_2(SU(N)/U(1)^{N-1}) = \Pi_1(U(1)^{N-1}) = Z^{N-1} .
  $
  In particular,
    $
  \Pi_2(SU(2)/U(1)) = \Pi_1(U(1)) = Z ,
  $
  see argument in Ref. \cite{KSW87}.
}
\begin{eqnarray}
  U(x)[ \partial_X , \partial_Y  ] U^\dagger(x)
  = - 2\pi n i \delta(X) \delta(Y) \theta(-Z) \sigma_3 .
  \label{sing}
\end{eqnarray}
To prove this, we first show that
\begin{eqnarray}
   [ \partial_X , \partial_Y  ] \varphi(x)
  = 2\pi n \delta(X) \delta(Y)  .
  \label{multi}
\end{eqnarray}
This is a result of the Stokes theorem; for the arbitrary
2-dimensional region $S$ including $(X,Y)=(0,0)$,
\begin{eqnarray}
   \int_S dX dY [ \partial_X , \partial_Y  ] \varphi
   = \int_S d^2S \det \pmatrix{\partial_X & \partial_Y \cr
   \partial_X \varphi & \partial_Y \varphi }
   = \int_S d^2S \nabla \times (\nabla \varphi)
   \nonumber\\
   = \oint_{C=\partial S} \partial_\mu \varphi dx^\mu 
   = \Delta \varphi
   = 2\pi n
  = 2\pi n \int_S dX dY \delta(X) \delta(Y)  ,
\end{eqnarray}
where the integer $n$ comes from the multi-valuedness of
$\varphi$.
\par
When $\theta=0$ (i.e., on the positive $Z$ axis),
\begin{eqnarray}
 U(x)_{0,\varphi}
=  \pmatrix{1 &  0 \cr   0 &   1}  ,
\end{eqnarray} 
which does not give non-trivial contribution in
(\ref{sing}). On the other hand, for $\theta=\pi$  (i.e.,
on the negative
$Z$ axis), 
\begin{eqnarray}
 U(x)_{\pi,\varphi}
= \pmatrix{0  
&  e^{+i\varphi(x)} \cr   - e^{-i\varphi(x)} &   0  } .
\end{eqnarray} 
Then, using (\ref{multi}), 
\begin{eqnarray}
   U(x)_{\pi,\varphi}[ \partial_X , \partial_Y  ]
U(x)_{\pi,\varphi}^\dagger
   = - i [ \partial_X , \partial_Y  ] \varphi(x) \sigma_3
  = - 2\pi i n \delta(X) \delta(Y) \sigma_3  .
\end{eqnarray}
This proves the statement (\ref{sing}).
\par
 The relation (\ref{sing}) shows that the term 
${i \over g}(U[\partial_\mu , \partial_\nu]
U^\dagger)^{(3)}$
 produces the magnetic field only along the negative $Z$
axis,
\begin{eqnarray}
 B_Z^{Ds}  
 := {i \over g} 
 (U[ \partial_X , \partial_Y  ] U^\dagger)^{(3)} 
 = {4\pi n \over g} \delta(X) \delta(Y) \theta(-Z)  .
\end{eqnarray}
So this is identified with the Dirac string (not
the magnetic monopole) extending from the origin to infinity
along the negative
$Z$ axis (due to the above choice of U) in
three-dimensional space.  Hence the divergence of
$B_Z^{Ds}$ is non-zero at the origin,
\begin{eqnarray}
 k_0^{Ds} = \nabla \cdot B_Z^{Ds} := \partial_Z   {i \over
g} 
 (U[ \partial_X , \partial_Y  ] U^\dagger)^{(3)}
 = - {4\pi n \over g} \delta^3(x) ,
\end{eqnarray}
which should be compared with (\ref{Ds}).

\par
Finally, we give an alternative definition of the
abelian-projected field strength,
\begin{eqnarray}
 f_{\mu\nu}
=   \tr(T^3 i g   
   [ {\cal A}^s_{\mu},  {\cal A}^s_{\nu}])
 + \tr(T^3 {i \over g}   
   U [\partial_\mu , \partial_\nu] U^\dagger) .
\end{eqnarray}
This is the abelian field strength obtained from the
singular gauge potential and consists of the magnetic
monopole part and the Dirac string part as shown above.
In the RHS, the second term 
$
\tr(T^3 {i \over g}   
   U [\partial_\mu , \partial_\nu] U^\dagger)
$
expresses the magnetic field on the Dirac
string and vanishes elsewhere.  
Therefore, the remaining part 
$
\tr(T^3 i g [ {\cal A}^s_{\mu},  {\cal A}^s_{\nu}])
$
denotes the field strength of the magnetic monopole
defined everywhere. Hence, the magnetic monopole part of
the magnetic current defined by
\begin{eqnarray}
 k_\rho := {1 \over 2}
 \epsilon_{\mu\nu\rho\sigma} \partial^{\sigma}
f_{\mu\nu}
\end{eqnarray}
is equivalent to  
\begin{eqnarray}
 K_\rho  = {1 \over 2}
  \epsilon_{\mu\nu\rho\sigma} \partial^{\sigma} 
   (g  \epsilon^{ab3} A_\mu^{a}   A_\nu^{b} )   .
\end{eqnarray}
In the three dimensional slices, this describes the
magnetic monopole with magnetic
charge 
\begin{eqnarray}
  g_m := \int K_0(x) d^3x = {4\pi \over g}n,
  \quad 
  {g g_m \over 4\pi} = n ,
\end{eqnarray}
where $n$ is an integer.
This is nothing but  the Dirac quantization condition.

\par
In the original YM theory, as a result of the Jacobi
identity,
\begin{eqnarray}
 \epsilon_{\mu\nu\rho\sigma} 
 [ {\cal D}_\nu, [ {\cal D}_\rho, {\cal D}_\sigma]] = 0 ,
\end{eqnarray}
the Bianchi identity always holds,
\begin{eqnarray}
 0 = \epsilon_{\mu\nu\rho\sigma} {\cal D}_\nu 
 {\cal F}_{\rho\sigma}, \quad
 \epsilon_{\mu\nu\rho\sigma} \partial_\nu 
 {\cal F}_{\rho\sigma}
 = ig \epsilon_{\mu\nu\rho\sigma} {\cal A}_\nu 
 {\cal F}_{\rho\sigma}
 = 2 ig {\cal A}_\nu \tilde {\cal F}_{\mu\nu} .
\end{eqnarray}
After gauge fixing, the Bianchi identity for the residual
U(1) is violated,
\begin{eqnarray}
 \epsilon_{\mu\nu\rho\sigma} \partial^{\rho} 
f_{\mu\nu}' 
 \not=  0 ,
\end{eqnarray}
which leads to the magnetic monopole.
In the original YM theory, the  magnetic monopole
does not exist.  However, note that
\begin{eqnarray}
 \epsilon_{\mu\nu\rho\sigma} \partial^{\rho} 
 {\cal F}_{\mu\nu}'^{(3)} \not= 0,
 \quad  {\cal F}_{\mu\nu}'^{(3)}
 := 2 \tr(T^3 {\cal F}_{\mu\nu}') ,
\end{eqnarray}
since  
\begin{eqnarray}
 \epsilon_{\mu\nu\rho\sigma} \partial^{\rho} 
{\cal F}_{\mu\nu}'^{(3)}
 &=&  \epsilon_{\mu\nu\rho\sigma} \partial^{\rho}
( \partial_\mu  a_\nu'  -   \partial_\nu   a_\mu' )
  - 
  \epsilon_{\mu\nu\rho\sigma} \partial^{\rho} 
  ig ([{\cal A}_\mu' ,  {\cal A}_\nu'  ])^{(3)} 
\nonumber\\
&=& \epsilon_{\mu\nu\rho\sigma} \partial^{\rho} 
  {i \over g} 
 (U[ \partial_\mu , \partial_\nu  ] U^\dagger)^{(3)} .
\end{eqnarray}
The RHS is equal to the Dirac string
contribution \cite{LRQ96,CG95}

\par
 Incidentally, the four-vector
\begin{eqnarray}
 K_\rho^{Ds}  = {1 \over 2} \epsilon_{\mu\nu\rho\sigma}
\partial^{\rho}   {i \over g} 
 (U[ \partial_\mu , \partial_\nu  ] U^\dagger)^{(3)} ,
\end{eqnarray}
denotes the trajectory
\begin{eqnarray}
 K_\mu(x)  =  \int d\tau {\partial y_\mu(\tau) 
 \over \partial \tau}
\delta^{(4)}(x-y(\tau,0)) ,
\end{eqnarray}
as the boundary $x_\mu=y_\mu(\tau, 0)$ of
the Dirac sheet described by $y_\mu(\tau, \sigma)$ (world
sheet of the Dirac string, i.e., 2-dimensional surface
swept by the Dirac string in 4-dimensional space),
\begin{eqnarray}
 \omega_{\mu\nu}(x) := {i \over g} 
 (U(x)[ \partial_\mu , \partial_\nu  ] U^\dagger(x))^{(3)}
 = \int d\tau d\sigma {\partial(y^\mu, y^\nu) \over
\partial(\tau, \sigma)} \delta^{(4)}(x-y(\tau,\sigma)) .
\end{eqnarray}


\newpage
\baselineskip 10pt
\begin{thebibliography}{99}
\bibitem{Nambu74}
  Y. Nambu,
  Strings, monopoles, and gauge fields,
  Phys. Rev. D 10, 4262-4268 (1974).

\bibitem{tHooft81}
  G. 't Hooft,
  Topology of the gauge condition and new confinement
phases in non-Abelian gauge theories,
  Nucl. Phys. B 190 [FS3], 455-478 (1981).
  
\bibitem{Mandelstam76}
  S. Mandelstam,
  Vortices and quark confinement in non-abelian gauge
theories, 
  Phys. Report, 23, 245-249 (1976).
  
\bibitem{Dirac31}  
  P.A.M. Dirac,
  Qantized singularities in the electromagnetic field,
  Proc. Roy. Soc., London, A 133, 60-72 (1931).
  \\
  The theory of magnetic poles,
  Phys. Rev.  74, 817-830 (1948).

\bibitem{tHooft78}
  G. 't Hooft,
  On the phase transition towards permanent quark
confinement,
  Nucl. Phys. B 138, 1-25 (1978).
  \\
  L. Del Debbio, M. Faber, J. Greensite and S. Olejnik,
    Center dominance, center vortices and confinement,
    hep-lat/9708023.
 \\
  M. Baker, J.S. Ball and F. Zachariasen,
  Dual QCD: a review,
  Phys. Rept. 209, 73-127 (1991).
  
\bibitem{KSW87}
  A. Kronfeld, G. Schierholz and U.-J. Wiese,
  Topology and dynamics of the confinement mechanism,
  Nucl. Phys. B 293, 461-478 (1987).
  
\bibitem{KLSW87}
  A. Kronfeld, M. Laursen, G. Schierholz and U.-J. Wiese,
  Monopole condensation and color confinement,
  Phys. Lett. B 198, 516-520 (1987).
  
\bibitem{Polikarpov96}
  M.I. Polikarpov,
  Recent results on the abelian projection of lattice
gluodynamics,
  hep-lat/9609020.
  \\
  M.N. Chernodub and M.I. Polikarpov,
  Abelian projections and monopoles,
  hep-th/9710205.

\bibitem{EI82}
  Z.F. Ezawa and A. Iwazaki,
  Abelian dominance and quark confinement in Yang-Mills
theories,
  Phys. Rev. D 25, 2681-2689 (1982).
  \\
  Abelian dominance and quark confinement in Yang-Mills
theories,
  II. Oblique confinement and $\eta'$ mass,
  Phys. Rev. D 26, 631-647 (1982).

\bibitem{Suzuki88}
  T. Suzuki, 
  A Ginzburg-Landau type theory of quark confinement,
  Prog. Theor. Phys. 80, 929-934 (1988).
  \\
  S. Maedan and T. Suzuki,
  An infrared effective theory of quark confinement based
on monopole condensation,
  Prog. Theor. Phys. 81, 229-240 (1989). 
  \\
  T. Suzuki, 
  Abelian confinement me+chanism in QCD,
  Prog. Theor. Phys. 81, 752-757 (1989).

\bibitem{SS91}
  J. Smit and A.J. van der Sijs,
  A magnetic monopole in pure SU(2) gauge theory,
  hep-lat/9312087,
  \\
  Monopoles and confinement,
  Nucl. Phys. B 335, 603-648 (1991).

\bibitem{SY90}
  T. Suzuki and I. Yotsuyanagi,
  Possible evidence of abelian dominance in quark
confinement,
  Phys. Rev. D 42, 4257-4260 (1990).
  \\
  S. Hioki, S. Kitahara, S. Kiura, Y. Matsubara, O.
Miyamura, S. Ohno and T. Suzuki,
  Abelian dominance in SU(2) color confinement,
  Phys. Lett. B 272, 326-332 (1991).
  \\
  Monopole distribution in momentum space in SU(2) lattice
gauge theory,
  Phys. Lett. B 285, 343-346 (1992).

\bibitem{BOT96}
  R.C. Brower, K.N. Orginos and C.-I. Tan,
  Magnetic monopole loop for the Yang-Mills instanton,
  hep-th/9610101,
  Phys. Rev. D 55, 6313-6326 (1997).
  Instantons in the maximally abelian gauge,
  hep-lat/9608012.
\\
  R.C. Brower, T.L. Ivanenko, J.W. Negele, K.N. Orginos,
  Instanton distribution in quenched and full QCD,
  hep-lat/9608086,
  Nucl. Phys. Proc. Suppl. 53, 547-549 (1997).

\bibitem{Shabanov97}
  S.V. Shabanov,
  The monopole dominance in QCD,
  hep-th/9611228.
  \\
  Abelian projection and studies of gauge-invariant
quantities in the lattice QCD without gauge fixing,
  hep-lat/9611029, 
  Mod. Phys. Lett. A11, 1081-1093 (1996). 

\bibitem{DT80}
  T.A. DeGrand and D. Toussaint,
  Topological excitations and Monte Carlo simulation of
Abelian gauge theory,
  Phys. Rev. D 22, 2478-2489 (1980).
  
\bibitem{Wilson} 
  K.G. Wilson,
  {\it Problems in physics with many scales of length},
  Sci. Am. 241, 158-179  (1979).
  {\it The renormalization group and critical phenomena},
  Rev. Mod. Phys. 55, 583-600 (1983).
  
\bibitem{SW94}
  N. Seiberg and E. Witten,
  Electric-magnetic duality, monopole condensation, and
confinement in N=2 supersymmetric Yang-Mills theory,
hep-th/9407087,
Nucl. Phys. B 426, 19 (1994).
\\
 Monopoles, duality and chiral symmetry breaking in N=2
supersymmetric QCD,
hep-th/9408099,
Nucl. Phys. B 431, 484 (1994).


\bibitem{CDG78}
  C.G. Callan, Jr., R. Dashen and D.J. Gross,
  Towards a theory of the strong interactions,
  Phys. Rev. D 17, 2717-2763 (1978).


\bibitem{SST95}
  S. Sasaki, H. Suganuma and H. Toki,
  Dual Ginzburg-Landau theory with QCD-monopoles for
dynamical chiral-symmetry breaking,
  Prog. Theor. Phys. 94, 373-384 (1995).
\\
  H. Suganuma, S. Sasaki and H. Toki,
  Color confinement, quark pair creation and dynamical
chiral-symmetry breaking in the dual Ginzburg-Landau theory,
  Nucl. Phys. B 435, 207-240 (1995).
\\
  S. Umisedo, S. Sasaki, H. Suganuma and H. Toki,
  Monopole dominance for dynamical chiral-symmetry breaking
in the dual Ginzburg-Landau theory,
  hep-ph/9609499.

\bibitem{Miyamura95}
  O. Miyamura,
  Chiral symmetry breaking in gauge fields dominated by
monopoles on SU(2) lattices,
  Phys. Lett. B 353, 91-95 (1995),
  
\bibitem{Halpern77}
  M.B. Halpern,
  Field-strength formulation of quantum chromodynamics,
  Phys. Rev. D 16, 1798-1801 (1977).



\bibitem{BFYM}
  M. Martellini and M. Zeni, 
  Feynman rules and $\beta$-function for the BF Yang-Mills
theory,
  hep-th/9702035.
\\
  F. Fucito, M. Martellini and M. Zeni,
  The BF formalism for QCD and quark confinement,
  hep-th/9605018.
\\
  A.S. Cattaneo, P. Cotta-Ramusino, F. Fucito, M.
Martellini, M. Rinaldi, A. Tanzini and M. Zeni,
  Four-dimensional Yang-Mills theory as a deformation of
topological BF theory,
  hep-th/9705123.

\bibitem{QR97}
  M. Quandt and H. Reinhardt,
  Field strength formulation of SU(2) Yang-Mills theory in
the maximal abelian gauge: perturbation theory,
  hep-th/9707185.

\bibitem{MLP85}
  H. Min, T. Lee and P.Y. Pac,
  Renormalization of Yang-Mills theory in the abelian gauge,
  Phys. Rev. D 32, 440-449 (1985).

\bibitem{KU82}
  T. Kugo and S. Uehara,
  General procedure of gauge fixing based on BRS
invariance principle,
  Nucl. Phys. B 197, 378-384 (1982).
  
\bibitem{HN93}
  H. Hata and I. Niigata,
  Color confinement, abelian gauge and renormalization group
flow,
  Nucl. Phys. B 389, 133-152 (1993).

\bibitem{DPY84}
  D.I. Dyakonov, V.Yu. Petrov and A.V. Yung,
  Quasiclassical expansion in an external Yang-Mills field
and the approximate calculation of functional determinants,
  Sov. J. Nucl. Phys. 39, 150-155 (1984).
  Quasiclassical expansion of Yang-Mills heat kernels and
approximate calculation of functional determinants,
  Phys. Lett. B 130, 385-388 (1983).

\bibitem{Zwanziger71}
  D. Zwanziger,
  Local-lagrangian quantum field theory of electric and
magnetic charges,
  Phys. Rev. D 3, 880-891 (1971).
\\
  R.A. Brandt, F. Neri and D. Zwanziger,
  Lorentz invariance from classical particle paths in
quantum field theory of electric and magnetic charge,
  Phys. Rev. D 19, 1153-1167 (1979).

\bibitem{DSuper}
  J. Fr\"ohlich and P.A. Marchetti,
  Soliton quantization in lattice gauge theories,
  Commun. Math. Phys. 112, 343-383 (1987).
  \\
  L. Del Debbio, A. Di Giacomo and G. Paffuti,
  Detecting dual superconductivity in the ground state of
gauge theory,
 hep-lat/9403013,
  Phys. Lett. B 349, 513-518 (1995).
\\
  A. Di Giacomo, 
  Mechanisms for color confinement,
  hep-th/9603029.
\\
  R.W. Haymaker,
  Dual Abrikosov vortices in U(1) and SU(2) lattice gauge 
theories, 
hep-lat/9510035.

\bibitem{SS94}
  H. Shiba and T. Suzuki,
  Monopole action and condensation in SU(2) QCD,
  hep-lat/9408004,
  Phys. Lett. B 351, 519-527 (1995).
\\
  S. Kitahara, Y. Matsubara and T. Suzuki,
  Deconfinement transition and monopoles in $T\not=0$
SU(2) QCD, hep-lat/9411036,
  Prog. Theor. Phys. 93, 1-17 (1995).

\bibitem{Kondo95}
  K.-I. Kondo,
  Bosonization and duality of massive Thirring model,
  Prog. Theor. Phys. 94, 899-914 (1995).
  \\
  Thirring model as a gauge theory,
  Nucl. Phys. B 450, 251-266 (1995).

\bibitem{ST78}
  M. Stone and P.R. Thomas,
  Condensed monopoles and abelian confinement,
  Phys. Rev. Lett. 41, 351-353 (1978).
  
\bibitem{BS78}
  K. Bardakci and S. Samuel,
  Local field theory for solitons,
  Phys. Rev. D 18, 2849-2860 (1978).
  K. Bardakci,  
  Local field theory for solitons. II,
  Phys. Rev. D 19, 2357-2366 (1979).

\bibitem{NO73}
  H.B. Nielsen and P. Olesen,
  Vortex-line models for dual strings,
  Nucl. Phys. B 61, 45-61 (1973).

\bibitem{BMK77}
  T. Banks, R. Myerson and J. Kogut,
  Phase transitions in Abelian lattice gauge theories,
  Nucl. Phys. B 129, 493-510 (1977).
  
\bibitem{Polyakov77}
  A.M. Polyakov,
  Quark confinement and topology of gauge theories,
  Nucl. Phys. B 120, 429-458 (1977).

\bibitem{Rajaraman}
  R. Rajaraman,
  An introduction to solitons and instantons in quantum
field theory 
  (Elsevier, Amsterdam, 1987).
  
\bibitem{SS96}
  T. Sch\"afer and E.V. Shuryak,
  Instantons in QCD,
  hep-ph/9610451.


  


\bibitem{CG95}
  M.N. Chernodub and F.V. Gubarev,
  Instantons and monopoles in maximal abelian projection of
SU(2) gluodynamics,
  hep-th/9506026.
  
\bibitem{HT96}
  A. Hart and M. Teper,
  Instantons and monopoles in the maximally abelian gauge,
  hep-lat/9511016,
  Phys. Lett. B371, 261-269 (1996).
 
\bibitem{BS96}
  V. Bornyakov and G. Schierholz,
  Instantons or monopoles? Dyons,
  hep-lat/9605019.
  
\bibitem{STSM95}
  H. Suganuma, A. Tanaka, S. Sasaki and O. Miyamura,
  Evidence of strong correlation between instanton and
QCD-monopole on SU(2) lattice,
  hep-lat/9512024.
  
\bibitem{FSST97}
  M. Fukushima, S. Sasaki, H. Suganuma, A. Tanaka, H.
Toki and D. Diakonov,
  hep-lat/9608084,
  Phys. Lett. B 399, 141-147 (1997).

\bibitem{FMT97}
  M. Feurstein, H. Markum and S. Thurner,
  Coexistence of monopoles and instantons for different
topological charge definitions and lattice actions,
  hep-lat/9702004.
  \\
  M. Feurstein, H. Markum and S. Thurner,
  Instantons and monopoles in lattice QCD.
  hep-lat/9702006.

\bibitem{LRQ96}
  K. Langfeld, H. Reinhardt and M. Quandt,
  Monopole and strings in Yang-Mills theories,
  hep-th/9610213.
  
\bibitem{KMST97}
  K.-I. Kondo, T. Murakami, T. Shinohara and Y. Taira,
  in preparation.

\bibitem{CD77}
  C.G. Callan, Jr., and R. Dashen,
  Pseudoparticles and massless fermions in two dimensions,
  Phys. Rev. D 8, 2526-2534 (1977).

\bibitem{RU78}
  S. Raby and A. Ukawa,
  Instantons in (1+1)-dimensional Abelian gauge theories,
  Phys. Rev. D 18, 1154-1173 (1978).
  
\bibitem{Levine80}
  H. Levine,
  Two-dimensional SU(N) Higgs theory, an instanton
approach,
  Nucl. Phys. B 170 [FS1], 128-138 (1980).
  
\bibitem{SI97}
  H. Suganuma and H. Ichie, private communications.

\bibitem{WY75}
  T.T. Wu and C.N. Yang,
  Concept of nonintegrable phase factors and glocal
formulation of gauge fields, 
  Phys. Rev. D 12, 3845-3857 (1975).
  \\
  Dirac monopole without strings: monopole harmonics,
  Nucl. Phys. B 107, 365 (1976).
  




\end{thebibliography}
\end{document}